\documentclass[%
 aip,
 amsmath,amssymb,
 reprint,%
]{revtex4-1}
\usepackage{graphicx}
\usepackage{dcolumn}
\usepackage{bm}
\usepackage[utf8]{inputenc}
\usepackage{booktabs} 
\usepackage{siunitx}  
\usepackage[T1]{fontenc}
\usepackage{mathptmx}
\usepackage{amsmath}
\usepackage{multirow}
\usepackage{subcaption}
\usepackage{caption}
\usepackage{etoolbox}
\usepackage{esdiff}
\usepackage{url}
\usepackage[justification=centering, singlelinecheck=false]{caption}
\usepackage{tikz}
\usetikzlibrary{shapes.geometric, arrows, arrows.meta, positioning,calc}
\makeatletter
\def\@email#1#2{%
 \endgroup
 \patchcmd{\titleblock@produce}
  {\frontmatter@RRAPformat}
  {\frontmatter@RRAPformat{\produce@RRAP{*#1\href{mailto:#2}{#2}}}\frontmatter@RRAPformat}
  {}{}
}%
\makeatother
\begin{document}

\preprint{AIP/123-QED}

\title[Unified Computational Approach to Variational Physics]{From Fermat's Principle to Physics Informed Neural Networks: A Unified Computational Approach to Variational Physics}
\author{Aman Razdan}
 \author{Aditya Shankar Mazumdar}
 \author{Amit Tanwar}
\author{Pragati Ashdhir}%
 \email{pragatiashdhir@hinducollege.ac.in}
\affiliation{ 
Department of Physics, Hindu College, University of Delhi, Delhi-110007, India
}%


\begin{abstract}
Variational principles form a unifying mathematical framework across many areas of physics, yet their instruction at the undergraduate level remains primarily analytical. This work presents a pedagogically oriented and computationally enhanced approach to variational modeling that integrates contemporary tools including gradient descent, automatic differentiation, and Physics-Informed Neural Networks (PINNs). Classical variational problems are reformulated as optimization tasks and implemented using open-source Python libraries such as NumPy, Matplotlib, PyTorch, and JAX. The proposed approach is demonstrated through a progression of problems drawn from standard undergraduate curricula, including the derivation of Snell’s law from Fermat’s principle, projectile motion with and without viscous drag, simple harmonic motion, nonlinear pendula with damping, steady-state heat conduction governed by the Laplace and Poisson equations with nonlinear temperature-dependent internal heat generation, the double pendulum via the principle of least action, and variational treatments of vibrating strings. In addition, quantum mechanical applications are presented through variational solutions of the hydrogen atom, helium atom, and a schematic nuclear model of the silicon nucleus, illustrating the breadth of the framework across classical, quantum, and nuclear physics. The approach aims to enhance conceptual understanding while simultaneously introducing students to modern computational research methodologies.
\end{abstract}

\maketitle

\section{\label{sec:level1}INTRODUCTION}

Variational ideas occupy a unifying role across diverse branches of physics, yet their pedagogical potential remains underutilized in undergraduate instruction. From the geometric principles underlying optical phenomena to contemporary applications involving deep neural networks, both nature and computation reveal a common organizing theme: physical behaviour is governed by extrema principles and the optimization of functionals. This shared structure provides a powerful opportunity for teaching—variational modeling can function not only as a conceptual bridge across traditionally separated subfields, but also as a practical and computationally accessible framework for solving physics problems.

Historically, variational reasoning entered physics with Fermat’s seventeenth-century assertion that light follows the path of least time\cite{tannery1891oeuvres,shapiro1973kinematic}. When placed on a rigorous mathematical foundation by Euler and Lagrange, this idea led to the Euler–Lagrange equations and enabled the modern formulation of constrained dynamics. Hamilton’s principle of stationary action and Noether’s theorem subsequently elevated variational structure to a central philosophical position within analytical mechanics, revealing that physical laws emerge from the extremization of appropriate functionals\cite{bailey1987dynamics}. Although these concepts are often encountered only piecemeal in undergraduate curricula, taken together they vividly demonstrate the unifying power of the variational viewpoint.

Advances in computational physics further expanded the reach of variational methods. Discrete extremization procedures—exemplified by the Ritz and Galerkin approaches and, later, the finite-element method—enabled differential equations to be solved approximately while preserving their underlying physical structure \cite{hulbert2010finite}. In this way, optimization evolved from a purely theoretical construct into a practical computational tool that is now integral to the modeling of real physical systems.

With the advent of automatic differentiation, variational computation acquired a new algorithmic foundation. By mechanizing the chain rule, these techniques deliver machine-precision derivatives even for large, high-dimensional models\cite{baydin2018automatic,griewank2008evaluating}. Reverse-mode differentiation—better known in physics education as backpropagation—restores a familiar structure: extremizing a functional by descending along its gradient. In this sense, deep-learning optimization is not a departure from physical reasoning but a modern extension of the calculus of variations.

Physics-Informed Neural Networks (PINNs) represent a contemporary embodiment of this convergence. By embedding governing differential equations directly into a loss functional, PINNs unite optimization, differentiation, and physical law within a single variational framework\cite{raissi2019physics}. For undergraduate learners, this provides a powerful conceptual message: machine-learning models need not be purely empirical, but can be grounded in the same physical principles encountered in introductory physics courses.

In this work, we bring these developments into an undergraduate-accessible framework using open-source Python-based libraries, NumPy and Matplotlib for numerical computation and visualization, PyTorch and JAX for automatic differentiation and neural network-based modeling. The framework is demonstrated through a structured progression of problems frequently encountered in undergraduate physics curricula. The elementary-level examples encompass Snell’s law of refraction derived from Fermat’s principle, projectile motion both with and without consideration of viscous drag, and the analysis of nonlinear pendulum. The intermediate-level illustrations include steady state heat conduction modeled via the Laplace and Poisson equations and the double pendulum examined through the principle of least action. Finally, the advanced-level problems employ the Rayleigh–Ritz method for the investigation of classical vibrating strings, the quantum mechanical treatment of the hydrogen and helium atoms, and nuclear structure modeling of the silicon nucleus utilizing the Woods–Saxon potential. These examples collectively illustrate the seamless integration of variational principles with sophisticated computational tools, offering students both profound conceptual clarity and valuable exposure to methodologies at the forefront of current research in computational physics.

\section{Methodological Framework}
This section develops the necessary background needed to understand the results in the paper and the analysis of the same.
\subsection{Theoretical Foundations}
We set the theoretical context of the work beginning with the mathematical introduction of functionals. While a standard function $f: x \to y$ maps a numeric input to a numeric output, a \textit{functional} $\mathcal{F}$ maps an entire function to a single scalar value:
\begin{equation}\mathcal{F}[f] = \lambda, \quad \text{where } \lambda \in \mathbb{R}
\end{equation}
Functionals are typically represented as definite integrals of the form
\begin{equation}
\mathcal{F}[f] = \int_a^b f(x)\,dx.
\end{equation}
Here, $\mathcal{F}$ takes the entire shape of the curve $f(x)$ as input and yields a single number. The central task of variational calculus is to find the specific function $f(x)$ that makes $\mathcal{F}$ stationary\cite{calcofvariations}. Analogous to finding a point $x_0$ where the derivative $f'(x_0)=0$ in ordinary calculus, we seek a function $f(x)$ (satisfying some boundary condition, say $f(a) = A$ and $f(b)=B$) such that the variation $\delta \mathcal{F} = 0$. Mathematically, we introduce a test path $\tilde{f}(x) = f(x) + \varepsilon g(x)$, where $g(x)$ is an arbitrary perturbation vanishing at the boundaries. The condition for stationarity requires that the functional's derivative with respect to the perturbation strength $\varepsilon$ vanishes, i.e.,
\begin{equation}\label{eq:calc_var}
\frac{d}{d\varepsilon}\mathcal{F}[f+\varepsilon g]\bigg|_{\varepsilon=0}=0
\end{equation}
This condition implies that $f(x)$ is a critical "point" in the infinite-dimensional space of functions. Nature appears to prefer these stationary paths. Many classical phenomena are governed by the minimization of specific fundamental functionals\cite{gelfand2000calculus}:
\begin{enumerate}
\item \textbf{Distance (Geodesics):} Objects in curved space follow the path that minimizes proper length.
\item \textbf{Time (Fermat's Principle):} Light propagates along the path that minimizes travel time, explaining the phenomena of reflection and refraction.
\item \textbf{Potential Energy (Dirichlet's Principle):} Static systems, such as soap films or hanging cables, settle into shapes that minimize total potential energy.
\item \textbf{Action (Hamilton's Principle):} Dynamical systems evolve along a trajectory that makes the Action stationary.
\end{enumerate}
This study focuses on the computational minimization of physically meaningful functionals, including optical time and mechanical action (Table~\ref{tab:fund_functionals}), to recover light trajectories, dynamical paths, and representative inverse variational problems encountered in undergraduate physics.
\begin{table}[!h]
\centering
\caption{Fundamental functionals minimized in this study. Here, $n(x,y)$ represents the refractive index, and $L(t,y,\dot{y})$ is the Lagrangian of the mechanical system.}
\renewcommand{\arraystretch}{2.0}
\setlength{\tabcolsep}{12pt}
\begin{ruledtabular}
\begin{tabular}{cc}
\textbf{Physical Quantity} &  \textbf{Functional Form} \\ \hline
Travel Time (Optics) &  $T[y]=\displaystyle{\int_{x_0}^{x_1} n(x,y)\sqrt{1+y'^2}\,dx}$ \\
Action (Mechanics) &   $S[y]=\displaystyle{\int_{t_0}^{t_1} L(t,y,\dot{y})\,dt}$ 
\end{tabular}
\end{ruledtabular}
\label{tab:fund_functionals}
\end{table}

\subsection{Computational Implementation}
Translating the variational ideas discussed in the preceding section into computational practice requires students to develop working knowledge of several complementary numerical methods and algorithms, which are outlined below.
\subsubsection{Automatic Differentiation}
\begin{figure}[!h]
\begin{tikzpicture}[scale=0.9, transform shape,
    main/.style={
        circle, 
        draw, 
        thick, 
        minimum size=0.8cm, 
        font=\bfseries, 
        inner sep=2pt
    },
    arrow/.style={
        -{Stealth[length=2.5mm]}, 
        shorten >=1pt, 
        shorten <=1pt
    },
    node distance=1.5cm and 2.5cm 
]

    
    \node[main] (x) {$u_1=x$};
    \node[main] (y) [below=of x] {$u_2=y$};
    
    \node[main] (a) [right=of x] {$u_3=a$};
    \node[main] (b) [right=of y] {$u_4=b$};
    
    \node[main] (z) at ($(a)!0.5!(b) + (2.5, 0)$) {$u_5=z$};

    
    \draw[arrow] (x) -- node[below, pos=0.9, font=\bfseries\Huge] {.} (a);
    
    \draw[arrow] (y) -- (a);
    
    \draw[arrow] (y) -- node[below, font=\bfseries] {sin} (b);
    
    \draw[arrow] (a) -- (z);
    
    \draw[arrow] (b) -- (z);

    \node[font=\bfseries] at ($(z.west) + (-0.25, 0)$) {+};

\end{tikzpicture}
\caption{Computational graph of $z$}
\label{fig:compgraph}
\end{figure}
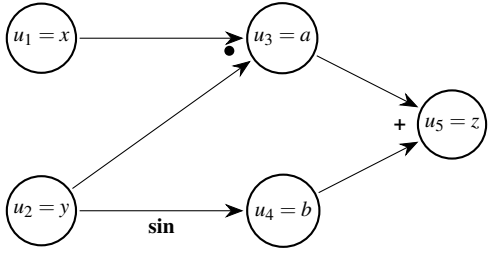

Automatic differentiation (AD) refers to a family of algorithmic techniques for computing derivatives of functions to machine precision. It achieves this by decomposing a complex computation into a sequence of elementary operations and systematically applying the chain rule at each step. Unlike symbolic differentiation, AD avoids the problem of rapidly growing expressions, and unlike finite-difference
approximations, it does not introduce truncation or round-off errors \cite{Admodes,bartholomew2000automatic}.\\\\
To illustrate the basic idea, consider the function
\begin{equation}
\label{eq:AD1}
z = f(x,y) = x\ y + \sin(y).
\end{equation}
Introducing intermediate variables
\begin{equation}
\label{eq:AD2}
a = x\ y, \qquad b = \sin(y),
\end{equation}
the computation may be written as
\begin{equation}
\label{eq:AD3}
z = a + b.
\end{equation}
This decomposition can be visualized in the form of directed acyclic graph, known as a
\emph{computational graph} shown in Fig.~\ref{fig:compgraph}. In this graphical depiction, the nodes represent variables and intermediate quantities, and edges represent elementary operations.

Within this framework, evaluating derivatives amounts to propagating
\emph{derivative information} through the computational graph using the chain
rule. Depending on the direction in which this propagation is carried out, two
distinct modes of automatic differentiation arise.

In \emph{forward-mode} AD, derivatives are propagated from the inputs toward the output, following the direction of the computation itself. Each node $u_i$ in the graph carries both its numerical value and its derivative with respect to a chosen independent variable $s$. If $u_j$ denotes
a child node of $u_i$, the chain rule takes the form
\begin{equation}
\label{eq:fwdAD}
\frac{\partial u_j}{\partial s}
=
\sum_i
\frac{\partial u_j}{\partial u_i}
\frac{\partial u_i}{\partial s}.
\end{equation}
As the computation proceeds, derivative information is accumulated alongside
the primal evaluation. In the example above, for $s=x$, we have
    \begin{eqnarray}
        \frac{\partial z}{\partial x}&&=\frac{\partial u_5}{\partial x}\nonumber\\
        &&=\frac{\partial u_5}{\partial u_3}\frac{\partial u_3}{\partial x}+\frac{\partial u_5}{\partial u_4}\frac{\partial u_4}{\partial x}\nonumber\\
        &&=\frac{\partial z}{\partial a}\frac{\partial a}{\partial x}+\frac{\partial z}{\partial b}\frac{\partial b}{\partial x}\nonumber\\
        &&=\frac{\partial a}{\partial x}+\frac{\partial b}{\partial x}
    \end{eqnarray} 
Thus computing $\partial z/\partial x$ requires tracking how variations in $x$ influence the intermediate variables $a$ and $b$, and subsequently how these variations propagate to $z$. We can perform a similar computation for $s=y$.

In \emph{reverse-mode} AD, the chain rule is implemented in the opposite direction. The computation first proceeds forward to evaluate and store all intermediate variables, thereby recording the computational trace. Derivatives are then propagated backward from the output toward the inputs. If $s$ denotes the scalar output and $u_i$ an intermediate node, the backward accumulation is given by
\begin{equation}
\label{eq:revAD}
\frac{\partial s}{\partial u_i}
=
\sum_j
\frac{\partial u_j}{\partial u_i}
\frac{\partial s}{\partial u_j},
\end{equation}
where the sum runs over all child nodes $u_j$ of $u_i$. For the present example,
the derivatives of $z$ with respect to both $x$ and $y$ are obtained
simultaneously in a single backward pass through the graph. We illustrate this below for $s=z$ and $i=1 \implies u_1=x$:
 \begin{eqnarray}
        \frac{\partial z}{\partial u_1}&&=\frac{\partial z}{\partial x}\nonumber\\
        &&=\frac{\partial z}{\partial u_3}\frac{\partial u_3}{\partial x}+\frac{\partial z}{\partial u_4}\frac{\partial u_4}{\partial x}\nonumber\\
        &&=\frac{\partial z}{\partial a}\frac{\partial a}{\partial x}+\frac{\partial z}{\partial b}\frac{\partial b}{\partial x}\nonumber\\
        &&=\frac{\partial a}{\partial x}+\frac{\partial b}{\partial x}
    \end{eqnarray}
Although the mathematical expressions in Eq.~\ref{eq:fwdAD} and  Eq.~\ref{eq:revAD} underlying forward and reverse modes appear formally similar, their computational behavior is fundamentally different. Forward mode computes derivatives alongside the primal evaluation and therefore requires relatively little memory, but it becomes inefficient when the number of input variables is large because each input requires a separate forward pass. Reverse mode, by contrast, incurs higher memory costs since the full forward trace must be stored, yet it enables the efficient computation of gradients with respect to all input variables in a single backward sweep. As a result, forward mode is best suited to problems with a small number of inputs and many outputs, while reverse mode is particularly effective for high-dimensional input spaces, as commonly encountered in variational formulations, inverse problems, and modern physics-based machine learning. 

In the present work, derivatives are evaluated using the
\texttt{jax.grad()} function from the \texttt{JAX} library\cite{jaxdocAD}, which provides an efficient implementation of reverse-mode automatic differentiation.

\subsubsection{Gradient Descent}
Gradient descent (GD) is an iterative optimization algorithm used to determine the
minimum of a function\cite{tapkir2023comprehensive}. For a differentiable scalar function $f(x)$, successive
approximations to the minimizing argument are obtained using the update rule
\begin{equation}
\label{eq:gd_scalar}
x_{i+1}
=
x_i
-
\eta
\left.
\frac{df}{dx}
\right|_{x=x_i},
\end{equation}
where $x_i$ and $x_{i+1}$ denote the values of the variable at the
$i^{\text{th}}$ and $(i+1)^{\text{th}}$ iterations, respectively, and
$\eta>0$ is the \emph{learning rate} (or step size), which controls the rate of
convergence of the algorithm.

For sufficiently small $\eta$, the iteration in
Eq.~(\ref{eq:gd_scalar}) converges toward a local minimum of $f(x)$, provided
$f$ is smooth and bounded from below. If $\eta$ is chosen too large, the
iterations may overshoot the minimum and fail to converge, while excessively
small values of $\eta$ lead to slow convergence.

In the present work, the objective is not a function of a single variable but a
\emph{functional}, whose argument is a discretized field. A continuous
functional of the form
\begin{equation}
\label{eq:functional}
\mathcal{F}[y]
=
\int_{a}^{b}
\mathcal{L}\big(y(x),y'(x),x\big)\,dx
\end{equation}
attains an extremum when the function $y(x)$ satisfies the Euler--Lagrange
equation
\begin{equation}
\label{eq:EL}
\frac{\partial \mathcal{L}}{\partial y}
-
\frac{d}{dx}
\left(
\frac{\partial \mathcal{L}}{\partial y'}
\right)
=0.
\end{equation}

Instead of solving Eq.~(\ref{eq:EL}) analytically, we discretize the domain and
approximate the functional $\mathcal{F}[y]$ by a scalar-valued multivariate
function
\begin{equation}
\mathcal{F}[y]
\;\longrightarrow\;
\mathcal{F}(y_1,y_2,\ldots,y_N),
\end{equation}
where $\mathbf{y}=(y_1,y_2,\ldots,y_N)$ represents the field values on a spatial
grid. In this discrete setting, the functional derivative is approximated by
the gradient of $\mathcal{F}$ with respect to $\mathbf{y}$.

The gradient descent update then generalizes naturally to
\begin{equation}
\label{eq:gd_functional}
\mathbf{y}_{i+1}
=
\mathbf{y}_i
-
\eta
\left.
\frac{\partial \mathcal{F}}{\partial \mathbf{y}}
\right|_{\mathbf{y}=\mathbf{y}_i},
\end{equation}
where $\mathbf{y}_i$ and $\mathbf{y}_{i+1}$ denote the arrays of field values at
the $i^{\text{th}}$ and $(i+1)^{\text{th}}$ iterations, respectively.

At convergence, the gradient
$\partial \mathcal{F}/\partial \mathbf{y}$ vanishes, and the discrete solution
approximates the stationary function that would be obtained from the
Euler--Lagrange equation in the continuum limit. In this sense, GD provides a computational realization of variational principles, transforming the problem of solving differential equations into one of iterative optimization.

This perspective aligns naturally with modern computational frameworks, where automatic differentiation enables efficient and accurate evaluation of gradients, making it possible to apply optimization algorithms directly to a wide class of problems in physics.

\subsubsection{Physics-Informed Neural Networks (PINNs)}
Physics-Informed Neural Networks (PINNs) represent a synthesis of deep learning
and physics-based modeling, in which neural networks are trained to satisfy the
governing equations of a physical system, rather than relying solely on labeled
data. In this framework, prior physical knowledge in the form of differential
equations, boundary conditions, and initial conditions is embedded directly
into the loss function of the network \cite{raissi2019physics}. The training
process is thus formulated as the minimization of a \emph{residual} (or loss)
functional, which maps a neural-network–parameterized function to a single
scalar value measuring the total error.

\begin{figure}[!h]
    \centering
    \includegraphics[width=\linewidth]{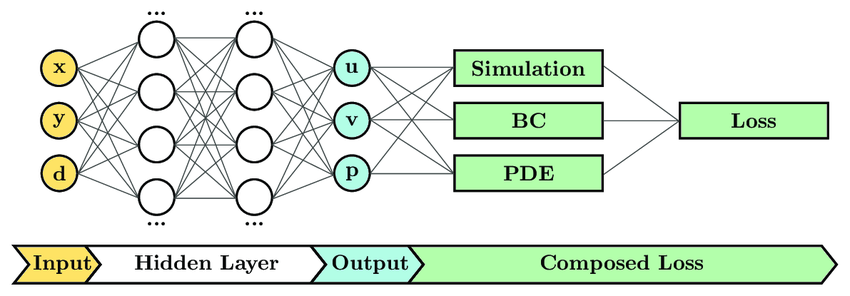}
    \caption{Schematic diagram of the physics-informed neural network architecture used for solving forward and inverse problems. The network takes spatial ($x, y$) and parameter coordinates ($d$) as input, passes them through hidden layers, and outputs state variables ($u, v, p$). The loss function is composed of data simulation loss, boundary condition (BC) loss, and PDE residual loss. Reprinted from Ref.~\onlinecite{pioch2023turbulence}}
    \label{fig:pinn_arch}
\end{figure}

Let the physical system be described either by an ordinary differential equation
(ODE) $\mathcal{N}[u(t)] = 0$ or by a partial differential equation (PDE) $\mathcal{N}[u(x,t)] = 0$, depending on the number of independent variables. In either case, the equation is defined over a domain $\Omega$, together with appropriate initial and/or
boundary conditions. In a PINN, the unknown field $u$ is approximated by a neural network $u_\theta$, where $\theta$ denotes the set of trainable parameters
(weights and biases). The neural network therefore plays the role of a flexible,
differentiable trial function (see Fig.~\ref{fig:pinn_arch}), capable of representing solutions to the
governing differential equation while being optimized to satisfy the associated
physical constraints.

The objective loss function $\mathcal{L}(\theta)$ is constructed to penalize
violations of both the governing equations and the imposed constraints. It is
typically expressed as a weighted sum of distinct scalar loss terms,
\begin{equation}
\label{eq:pinn_loss}
\mathcal{L}(\theta)
=
w_{\text{data}}\,\mathcal{L}_{\text{data}}
+
w_{\text{phys}}\,\mathcal{L}_{\text{phys}},
\end{equation}
or, equivalently, decomposed into contributions from the differential equation,
boundary conditions, and initial conditions,
\begin{equation}
\mathcal{L}(\theta)
=
\mathcal{L}_{\text{PDE}}
+
\mathcal{L}_{\text{BC}}
+
\mathcal{L}_{\text{IC}}.
\end{equation}

The data loss term $\mathcal{L}_{\text{data}}$ enforces fidelity to observed
measurements, boundary conditions, or initial conditions, and is commonly
evaluated as the mean squared error (MSE) between the network predictions and
available data points \cite{raissi2019physics,farea2024understanding}. The
physics loss term $\mathcal{L}_{\text{phys}}$ acts as a physics-based
regularizer. If the governing equation—whether an ordinary or partial
differential equation—can be expressed in residual form
$f(\boldsymbol{\xi}) \approx 0$, where $\boldsymbol{\xi}$ denotes the relevant
independent variables, the network employs automatic differentiation to
evaluate this residual at a set of scattered \emph{collocation points}
$\{\boldsymbol{\xi}_i\} \subset \Omega$,
\begin{equation}
\mathcal{L}_{\text{phys}}
=
\frac{1}{N_r}
\sum_{i=1}^{N_r}
\left|
\mathcal{N}[u_\theta(\boldsymbol{\xi}_i)]
\right|^2.
\end{equation}
This term penalizes the network whenever its predicted derivatives violate the
underlying physical law. Here, the collocation points are a set of user-chosen points in the domain at which the governing differential equation is enforced approximately, without introducing a spatial or temporal mesh.

Crucially, all derivatives--temporal, spatial, or both--appearing in
$\mathcal{N}[u_\theta]$ are computed using automatic differentiation, ensuring
machine-precision evaluation without resorting to numerical discretization. The
resulting optimization problem,
\begin{equation}
\label{eqn:loss}
\theta^{*} = \operatorname*{arg\,min_{\theta}} \mathcal{L}(\theta)
\end{equation}
is solved using gradient-based optimization algorithms such as gradient descent
or its modern variants. In Eq.~\ref{eqn:loss}, the operator {\texttt{\textbf{arg min}} returns the parameter values $\theta^{*}$ at which the loss attains its minimum, rather than the minimum value itself.

From a variational perspective, PINNs may be viewed as performing optimization
over a function space parameterized by neural networks. Instead of explicitly
solving the Euler--Lagrange equations associated with a functional, the network
parameters are iteratively updated so that the trial function $u_\theta(\boldsymbol{\xi}_i)$ approaches a stationary solution of the governing equations. In this sense, PINNs provide a computational bridge between classical variational principles and modern machine-learning–based optimization.

This formulation is particularly appealing for physics education and research,
as it unifies differential equations, variational methods, automatic
differentiation, and optimization within a single computational framework,
enabling a mesh-free approach to solving complex physical problems
\cite{seo2024solving}.

In our work, PINNs have been implemented using the \texttt{torch.nn} module (for initializing the neural network), the \texttt{torch.autograd} module (for computing automatic gradients), and the \texttt{torch.optim.Adam} class in the \texttt{torch.optim} module (as an optimizer) of the \texttt{PyTorch} library\cite{torchdocAD}.

\subsubsection{Algorithmic Procedure}
Since the proposed approach is applicable to a wide variety of physical systems, we outline a generic algorithmic framework that can be adapted to any problem by supplying the appropriate physical inputs. The flowchart shown in Fig.~\ref{F1} summarizes this procedure. For a given computational domain, which may be one- or multi-dimensional, we begin by initializing a trial field \(y\) (or path) with random values, subject to the prescribed boundary or initial conditions. Using this trial configuration, the corresponding functional \(S[y]\) is constructed according to the physical problem under consideration.

The gradient of the functional with respect to the field variables is then computed using automatic differentiation. This gradient information is employed to iteratively update the field by minimizing the functional using a gradient descent algorithm. The iterative procedure is continued until convergence is achieved or a predefined maximum number of iterations, denoted by \texttt{Max} or \texttt{Max\_Iter}, is reached. The parameter \(\eta\) represents the learning rate (or step size), which controls the magnitude of each update step and was introduced earlier in the context of gradient descent.
\begin{figure}[!h]
\centering
\begin{tikzpicture}[
    scale=0.95, transform shape, 
    node distance=1.4cm,
    startstop/.style={rectangle, rounded corners, minimum width=2cm, minimum height=0.8cm, text centered, draw=black, fill=red!10, font=\footnotesize},
    io/.style={trapezium, trapezium left angle=70, trapezium right angle=110, text width=2.8cm, minimum height=1cm, text centered, draw=black, fill=blue!10, inner sep=2pt, font=\footnotesize},
    process/.style={rectangle, text width=2.8cm, minimum height=0.9cm, text centered, draw=black, fill=orange!10, font=\footnotesize},
    decision/.style={diamond, aspect=2, text width=2cm, minimum height=0.8cm, text centered, draw=black, fill=green!10, inner sep=0pt, font=\footnotesize},
    arrow/.style={thick,->,>=stealth}
]

    \node (start) [startstop] {Start};
    \node (init) [io, below of=start] {Initialize Domain $x$ \& Field $y$};
    \node (func) [process, below of=init] {Define Functional $S[y]$};
    \node (jax) [process, below of=func] {Compute Grad \\ \texttt{g\_fn = grad(S)}};

    \node (loop) [decision, right=1.2cm of init] {Iter $<$ Max?};
    \node (calc) [process, below of=loop, yshift=-0.2cm] {Compute $\nabla S$};
    \node (mask) [process, below of=calc] {Mask B.C. \\ $0\leftarrow\nabla S_{bc}$};
    \node (update) [process, below of=mask] {Update \\ $y \leftarrow y - \eta \nabla S$};
    
    \node (stop) [startstop, above of=loop] {Return $y^*$};

    
    \draw [arrow] (start) -- (init);
    \draw [arrow] (init) -- (func);
    \draw [arrow] (func) -- (jax);

    \path (init.east) -- coordinate[midway] (gapcenter) (loop.west);
    \draw [arrow] (jax.east) -- (gapcenter |- jax.east) |- (loop.west);

    \draw [arrow] (loop.south) -- node[right, font=\scriptsize] {Yes} (calc.north);
    \draw [arrow] (calc) -- (mask);
    \draw [arrow] (mask) -- (update);

    \draw [arrow] (update.south) -- ++(0,-0.4) -| ($(loop.east) + (0.9,0)$) -- (loop.east);

    \draw [arrow] (loop.north) -- node[midway, right, font=\scriptsize] {No} (stop.south);

\end{tikzpicture}
\caption{Generic algorithmic workflow for the variational optimization framework used in this work. Starting from an initial trial field \(y\) defined over the computational domain \(x\), a problem-specific functional \(S[y]\) is constructed. The functional gradient \(\nabla S\) is evaluated using automatic differentiation and iteratively minimized using gradient descent with learning rate \(\eta\). Boundary conditions are enforced by masking the gradient at constrained points. The iteration continues until the prescribed maximum number of iterations is reached, yielding the optimized field \(y^\ast\).}
\label{F1}
\end{figure}
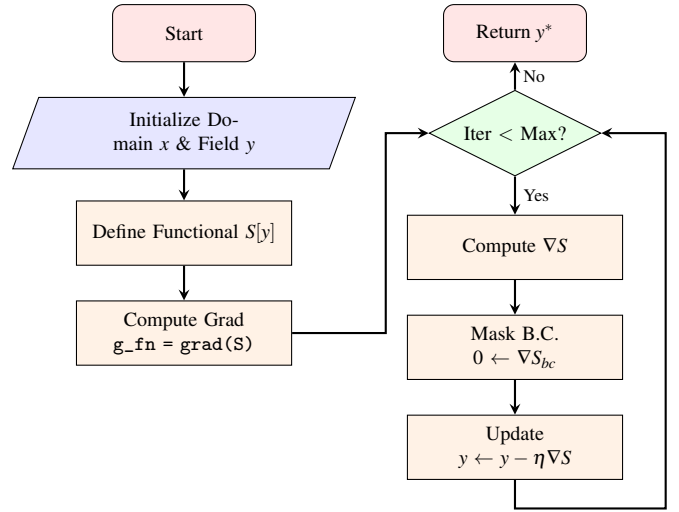

\section{Illustrative Examples}
We demonstrate the scope and effectiveness of the proposed computational framework through two complementary classes of problems: \emph{forward variational problems}, where the governing functional is known, and \emph{inverse variational problems}, where the functional is constructed from a prescribed differential equation. In each case, we focus on the underlying physical ideas and the principal results, while the details of the numerical implementation and the associated Python codes are provided as \emph{Chapters in the Supplementary Material}. The relevant chapter number is explicitly cited at the conclusion of each illustrative example.
\subsection{The Forward Variational Problems}
Here, we address systems where the governing functional is known explicitly and serves as the fundamental starting point. The integral defining the functional is discretized, and its minimum is obtained numerically through direct optimization.
\begin{figure*}
\centering
    \begin{subfigure}[b]{0.47\linewidth}
        \includegraphics[width=\linewidth]{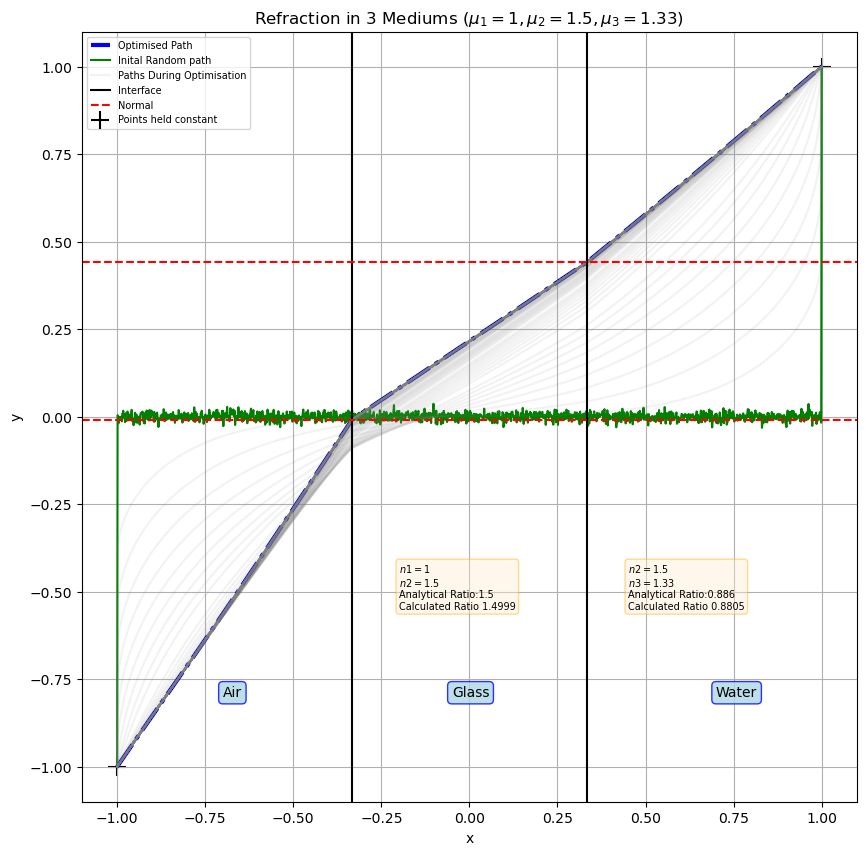}
        \caption{Case 1: Rarer-Denser-Rarer}
        \label{fig:rdr}
    \end{subfigure}
    \hfill
    \begin{subfigure}[b]{0.47\linewidth}
        \includegraphics[width=\linewidth]{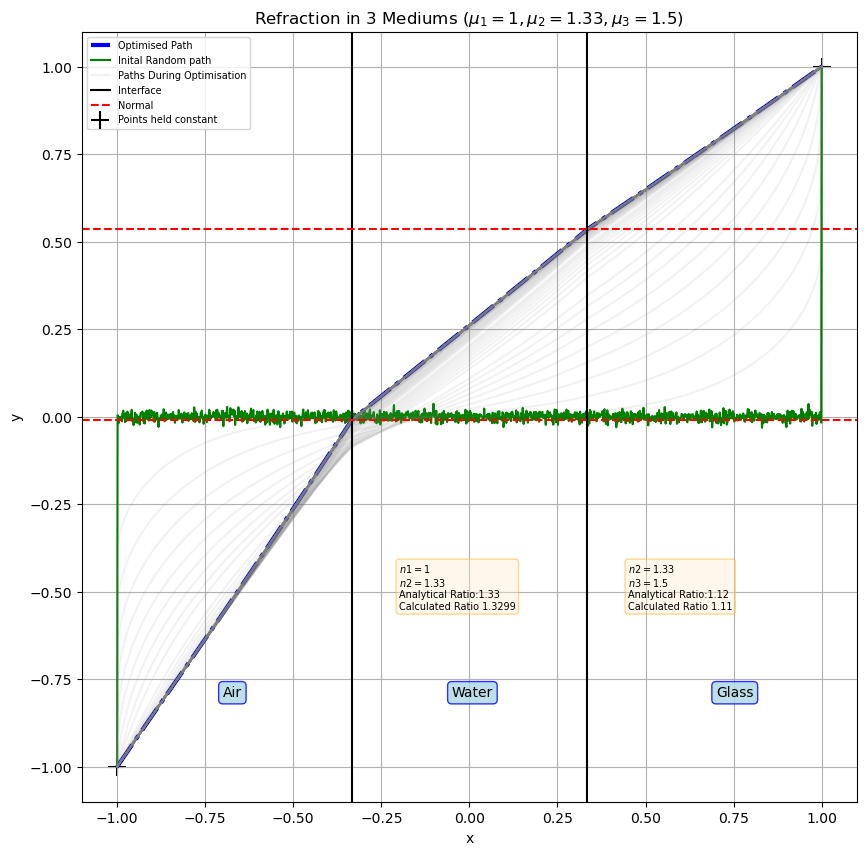} 
        \caption{Case 2: Rarer-Denser-Denser}
        \label{fig:rdd}
    \end{subfigure}
    \caption{\textbf{Refraction across three mediums}: For numerical convenience the value of c was taken to be 1 , the $x-domain$ was divided from $-1$ to $1$ into $1,500$ points,so each interface had 500 points. The initial path is assumed to be normally distributed around 0 with a variance of 1 with the end points fixed at $(-1,-1)$ and $(1,1)$ , giving us the pattern we can see in the image, }
    \label{fig:opt_system}
\end{figure*}
\subsubsection{Fermat’s Principle: Least time in Optics}
The fundamental functional in optics is the travel time. \emph{Fermat's principle} dictates that light follows the path that minimizes this time, yielding \emph{Snell's Law} of refraction:
\begin{equation}
n_1 \sin \theta_1 = n_2 \sin \theta_2,
\end{equation}
where $n_i$ represents the refractive index in the $i^{th}$ medium and $\theta_i$ the angle relative to the normal in that medium.
By discretizing the time functional (Table~\ref{tab:fund_functionals}) and applying gradient descent, we determine the optical path directly. The optimization successfully recovers the trajectory predicted by Snell's Law validating the computational framework for geometric optics.

Two distinct arrangements were considered within this study:
\begin{enumerate}
\item Transition from Air $\rightarrow$ Glass $\rightarrow$ Water (see Fig.~\ref{fig:rdr})
\item Transition from Air $\rightarrow$ Water $\rightarrow$ Glass (see Fig.~\ref{fig:rdd})
\end{enumerate}

To facilitate this investigation, we fixed two boundary points at the
coordinates $(-1,-1)$ and $(1,1)$. An $x$-array spanning this
interval was constructed and divided into three equal segments, with
each segment assigned its corresponding refractive index to represent
three distinct optical media. The light path between the fixed endpoints was represented by a discrete array $y(x)$. This array was initialized randomly, subject
to the constraint that the boundary values remained fixed at
$y(-1)=-1$ and $y(1)=1$. Such a random initialization ensures that
no prior geometric bias is imposed on the solution and allows the
optimization process to identify the extremal path solely from the
variational principle. Starting from this initial random path across the three media, the optical path length functional was minimized using the gradient
descent algorithm. Through successive iterations, the path evolved
towards a stationary solution that satisfies Fermat’s principle,
ultimately recovering the physically correct refracted trajectory. \\
To determine the angles of incidence and refraction and thereby verify
Snell’s law, we analyzed the optimized light path in the vicinity of
the material interfaces. The interfaces were configured to be parallel
to the $y$-axis, so that the corresponding surface normals were
aligned with the $x$-axis. Under this geometry, the angle between a
ray and the normal is identical to the angle between the ray and the $x$-axis. The ray angle within each medium was obtained numerically by computing
the local slope of the optimized path. Specifically, the slope was
evaluated using the discrete points adjacent to the interface: the
difference between the $y$-coordinate at the interface and the
nearest point within the corresponding medium was divided by the
uniform grid spacing $dx$. The \emph{arctangent} of this slope yields the corresponding ray angle. The angles extracted in this manner show excellent agreement with those predicted by analytical calculations (Table~\ref{tab:refraction_indice}), thereby confirming Snell’s law and validating the variational and computational approach. See Chapter-1 of \emph{Supplementary Material} for source codes.
\begin{table}[!h]
\centering
\caption{Comparison of relative refractive indices obtained from tabulated material constants 
(\(n_2/n_1\)) with those extracted from the numerically optimized ray path using 
Snell’s law, \(n_2/n_1 = \sin i / \sin r\).}
\renewcommand{\arraystretch}{1.6}
\setlength{\tabcolsep}{6pt}
\begin{ruledtabular}
\begin{tabular}{lcccc}
\textbf{Interface} &
\textbf{$i$ (°)} &
\textbf{$r$ (°)} &
\textbf{$n_2/n_1$} &
\textbf{$\sin i/\sin r$} \\ \hline
Air $\rightarrow$ Water   & 56.7479 & 38.95 & 1.3300 & 1.3303 \\
Water $\rightarrow$ Glass & 38.9500 & 34.35 & 1.1278 & 1.1141
\end{tabular}
\end{ruledtabular}
\label{tab:refraction_indice}
\end{table}

\subsubsection{Hamilton’s Principle: Stationary Action in Mechanics}
\emph{Hamilton’s principle} states that the actual motion of a physical system between two fixed times occurs along the path for which the action is stationary. A complementary and contemporary interpretation is emphasized by Greydanus \textit{et al.}\cite{greydanus2023nature}, who characterize the action as ``Nature’s cost function.'' From this viewpoint, physical evolution may be understood as an optimization process: among all admissible paths consistent with the boundary conditions, the realized trajectory extremizes a single scalar functional, the \emph{action} $S[y]$ defined in Table~\ref{tab:fund_functionals}. To highlight the significance of this interpretation and its instructional value, we work out three representative problems.
\subsubsection*{Projectile Motion}
The first is a simpler one-dimensional \emph{projectile motion}. Let the height of the projectile of mass $m$, as a function of time, be denoted by $y(t)$. Then, the Lagrangian of the system is defined as the difference of the kinetic ($T$) and potential energy ($V$):
\begin{equation}\label{eqn:proj_lagran}
\mathcal{L}(t,y,\dot{y})=T-V=\frac{1}{2}m\dot{y}^2-mgy,
\end{equation}
From Table.~\ref{tab:fund_functionals}, we know that action is then defined as
\begin{equation}\label{eqn:action_proj}
S[y(t)]=\int_{t_0}^{t_1}\mathcal{L}(t,y,\dot{y})\,dt=\frac{m}{2}\int_{t_0}^{t_1}(\dot{y}^2-2gy)\,dt
\end{equation}
By discretizing the integral in Eq.~(\ref{eqn:action_proj}) for an initial $[y_0]$ array and minimizing using gradient descent, we are able to determine the optimal trajectory, which maps with the analytical results (see Fig.~\ref{fig:proj_cons}).\\ Refer Chapter-2 Section 2.1 of \emph{Supplementary Material} for source codes.
\begin{figure}[!h]
\centering
    \begin{subfigure}[b]{0.49\linewidth}
        \includegraphics[width=\linewidth]{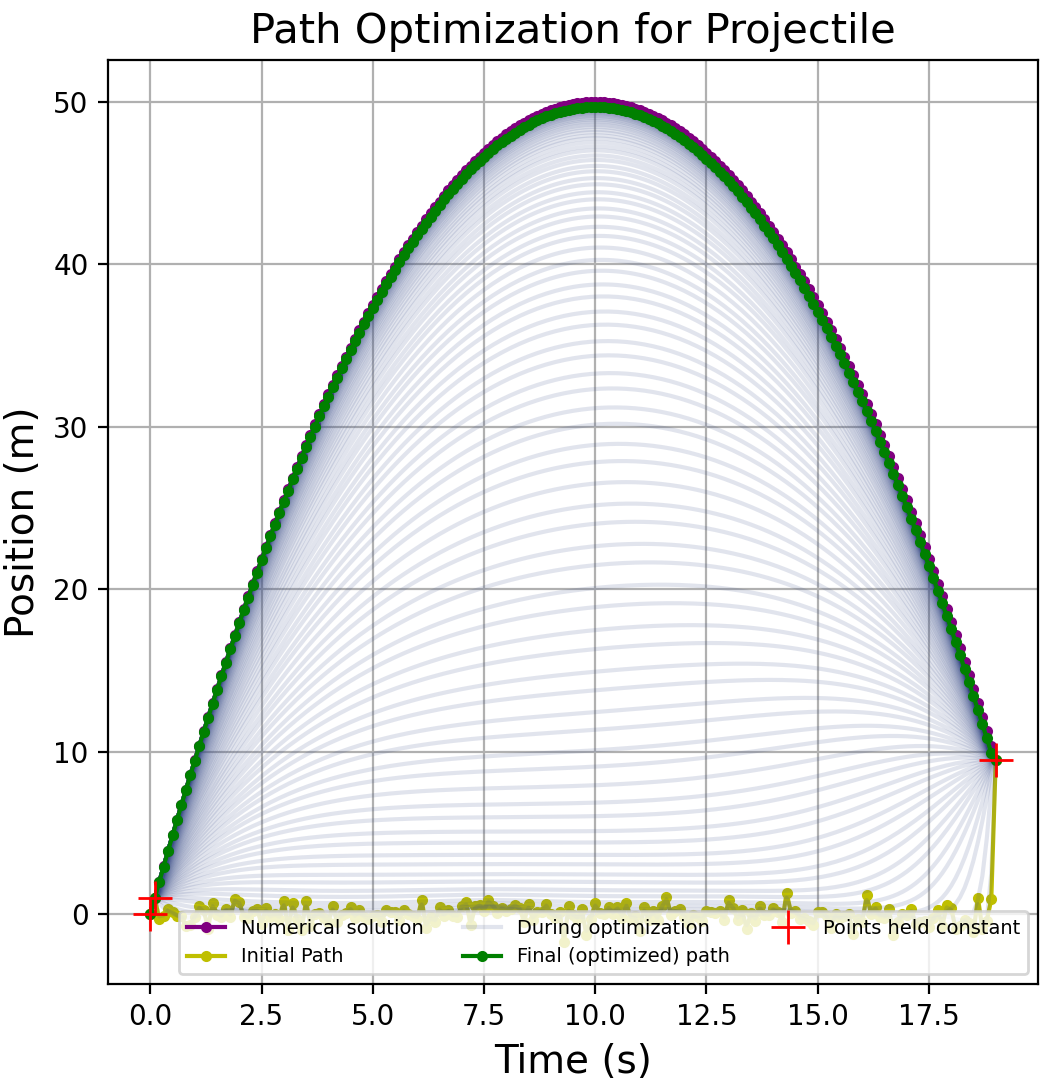}
        \caption{}
        \label{fig:proj_cons}
    \end{subfigure}
    \hfill
    \begin{subfigure}[b]{0.49\linewidth}
        \includegraphics[width=\linewidth]{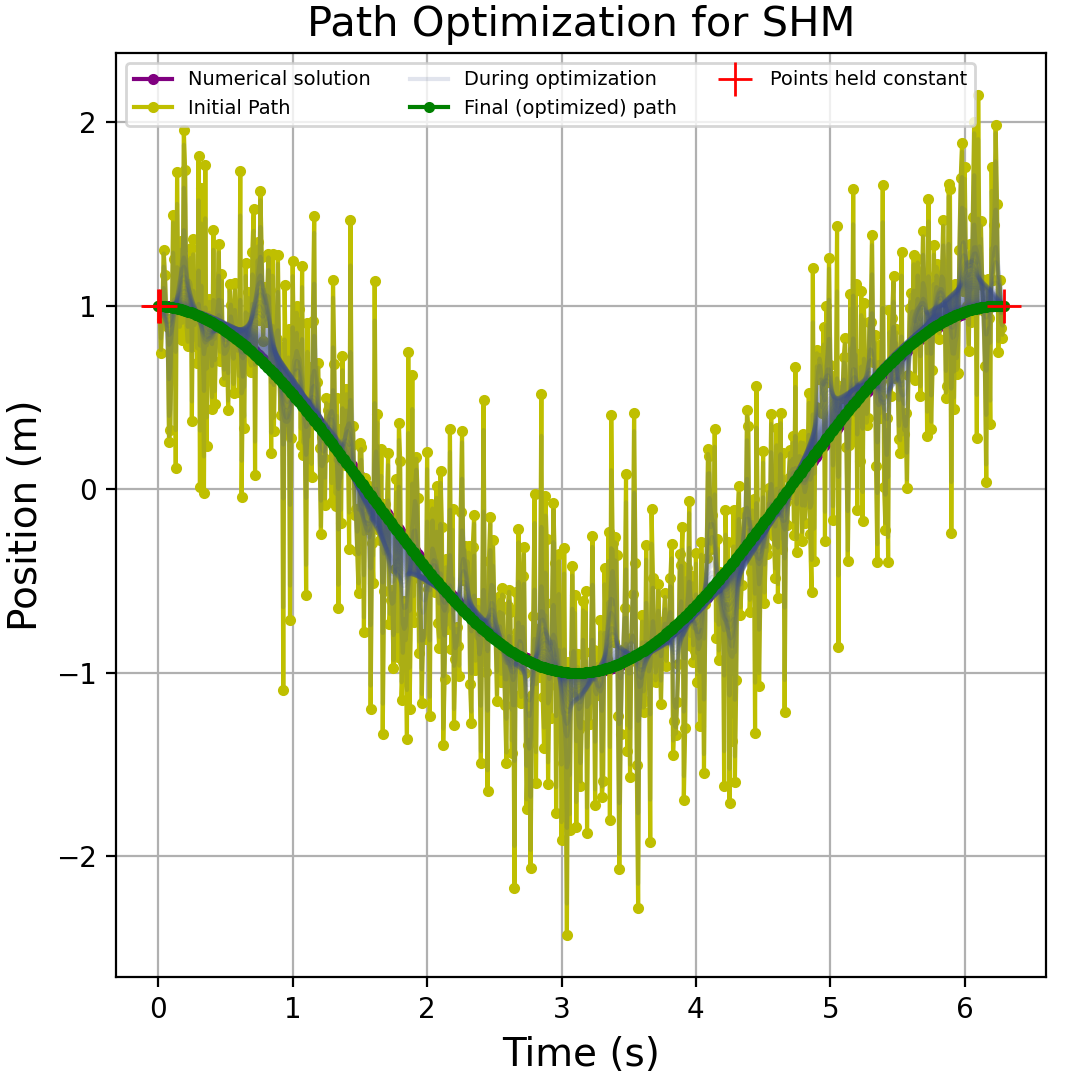} 
        \caption{}
        \label{fig:shm_cons}
    \end{subfigure}
    \caption{\textbf{Path optimization via action minimization}. Evolution of the trajectory for \textbf{(a)} projectile motion and \textbf{(b)} a simple harmonic oscillator. The optimizer relaxes the initial trial path (yellow) through intermediate states (faint blue) into the final stationary action path (green), which overlaps perfectly with the RK4 benchmark (purple). All physical constants were set to unity. Time steps were $dt=0.1$ s for the projectile ($t \in [0, 19]$ s) and $dt=0.01$ s for the oscillator ($t \in [0, 10]$ s).}
    \label{fig:mech_system}
\end{figure}
\subsubsection*{Simple Harmonic Motion}
The second problem focuses on the periodic motion of a \emph{simple harmonic oscillator (SHM)}, a basic model that underlies a wide range of physical systems. Let the displacement of a mass $m$, attached to a spring of force constant $k$, be $x(t)$ from its equilibrium position at some instant $t$. Then, for this system, we have the Lagrangian
\begin{equation}\label{eqn:shm_lagran}
    \mathcal{L}(t,x,\dot{x})=\frac{1}{2}m\dot{x}^2-\frac{1}{2}kx^2
\end{equation}
and, the action
\begin{equation}\label{eqn:action_shm}
    S[y(t)]=\int_{t_0}^{t_1}\mathcal{L}(t,x,\dot{x})\,dt=\frac{1}{2}\int_{t_0}^{t_1}(m\dot{x}^2-kx^2)\,dt
\end{equation}
But here, straightforward minimization of action does not yield the right results due to a slight oversight, which is referred to as the \emph{unconstrained energy effect}\cite{greydanus2023nature}. Basically, the periodic motion of the mass in this case can itself be viewed as a constraint of the motion, which needs to be encoded within the computation. We address this by enforcing energy conservation: we define an energy residual (the deviation of total energy from its initial value) and minimize a weighted combination of this residual and the action. If $w_s$ and $w_e$ are the weights associated with the action and the energy residual respectively, then the modified action is given as 
\begin{equation}\label{eqn:weightstuning}
S'=w_sS+w_e\Delta E.
\end{equation}
With tuned weights, this approach recovers the correct numerical solution (see Fig.~\ref{fig:shm_cons}). \\ Refer Chapter-2 Section 2.1 of \emph{Supplementary Material} for source codes.
\subsubsection*{Double Pendulum}
Building on the results of the previous problem, the third example is extended to a coupled-mass system: a \emph{planar double pendulum}. Two point masses $m_1$ and $m_2$ are connected by rigid, massless rods of lengths $l_1$ and $l_2$. Mass $m_1$ is attached to a fixed pivot by the first rod and makes an angle $\theta_1$ with the vertical, while mass $m_2$ is attached to $m_1$ by the second rod and makes an angle \(\theta_2\) with the vertical. The Lagrangian of the system can then be written as
\small
\begin{align}
    \mathcal{L}(t,\theta_1,\dot{\theta}_1,\theta_2,\dot{\theta}_2)=&\frac{1}{2}m_1l^2_1\dot{\theta}_1^2+\frac{1}{2}m_2\{l^2_1\dot{\theta}^2_1+l_2^2\dot{\theta}^2_2\nonumber\\
    &+2l_1l_2\dot{\theta}_1\dot{\theta}_2\cos(\theta_1-\theta_2)\}\nonumber\\
    &+m_1gl_1\cos\theta_1\nonumber\\
    &+m_2g\{l_1\cos\theta_1+l_2\cos\theta_2\}
\end{align}
\normalsize
\begin{figure}[!h]
    \centering
    \includegraphics[width=\linewidth]{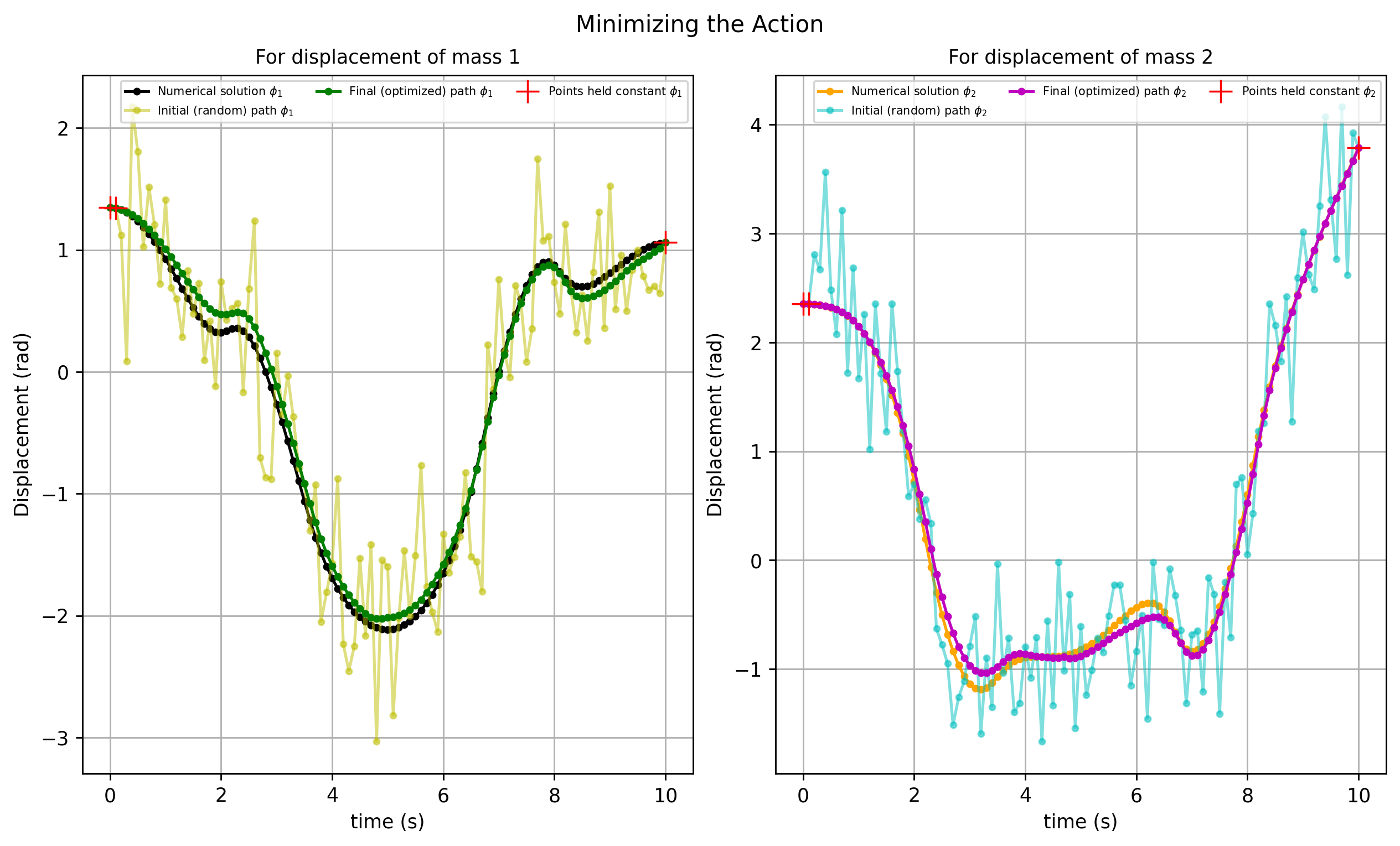}
    \caption{\textbf{Trajectory optimization for a double pendulum}. Displacement profiles for the first and second masses are shown. The optimizer relaxes the randomized initial paths (yellow and cyan) into the final stationary action trajectories (green and magenta), which align with the numerical solution (dashed lines). All physical parameters ($m, l, g$) were set to unity. The optimization used a time step of $dt=0.1$ s over the domain $t \in [0, 10]$ s}
    \label{fig:dblpend}
\end{figure}
The corresponding action is defined in the usual manner as the time integral of the Lagrangian. Since the system possesses two degrees of freedom, corresponding to the two angular coordinates, a weighted combination of the energy residual (as in the previous problem) and the action is minimized with respect to both angular displacements, $\theta_1$ and $\theta_2$. This variational optimization yields stationary-action trajectories that are in good agreement with the solutions obtained by direct numerical integration. (see Fig.~\ref{fig:dblpend}). Refer Chapter-2 Section 2.2 of \emph{Supplementary Material} for source codes.

\subsection{The Inverse Variational Problems} 
In many physical and mathematical settings, a system is described directly by a governing differential equation rather than by an explicit variational principle. In such cases, one may invoke the \emph{inverse problem of the calculus of variations}, wherein a suitable functional is deliberately constructed so that its Euler--Lagrange equation reproduces the given differential equation. Within this framework, the differential equation is treated as a constraint that guides the design of a \emph{synthetic action functional}. Once this functional is defined, the original boundary- or initial-value problem is recast as an optimization task: the desired solution is obtained by minimizing the functional over the space of admissible functions. The variational minimum corresponds to a function whose Euler–Lagrange condition enforces the target differential equation.

This perspective allows variational optimization to be repurposed as a general and flexible differential equation solver. In particular, it enables the use of gradient-based optimization, automatic differentiation, and machine-learning-inspired tools to solve ordinary and partial differential equations, even when no underlying physical action principle is known \emph{a priori}. From a pedagogical point of view, this approach bridges traditional differential equation methods and modern computational optimization, reinforcing the unifying role of variational thinking across physics and applied mathematics.
\subsubsection{Eigenvalue Problems: The Rayleigh-Ritz Functional}
A broad class of physical boundary value problems is encompassed by Sturm--Liouville theory~\cite{lutzen1984sturm}. The general Sturm--Liouville differential equation is
\begin{equation}
\label{eqn:SLDE}
\frac{d}{dx}\!\left[p(x)\frac{dy}{dx}\right]
+\bigl(q(x)+\lambda\,w(x)\bigr)y=0,
\end{equation}
where $p(x)$, $q(x)$, and the weight function $w(x)$ are known real-valued
functions defined on a closed interval $[a,b]$. The objective is to determine
the eigenfunction $y(x)$ and the corresponding eigenvalue $\lambda$ that satisfy
the prescribed boundary conditions.

Equation~\eqref{eqn:SLDE} arises as the \emph{Euler--Lagrange equation} associated with the stationary value of the functional~\cite{arfken2011mathematical}
\begin{equation}
\label{eqn:ritzfunc}
J[y]=\int_{a}^{b}\!\left[p(x)\left(\frac{dy}{dx}\right)^{2}
- q(x)y^{2}\right]dx,
\end{equation}
subject to the normalization constraint
\begin{equation}
\int_{a}^{b} w(x)\ y^{2}\,dx = 1.
\end{equation}
Equivalently, the problem may be formulated as the unconstrained minimization
of the Rayleigh quotient~\cite{basilevsky2013applied}
\begin{equation}
\label{eqn:rayl_quot}
\mathcal{F}[y]
=
\frac{\displaystyle\int_{a}^{b}
\left[p(x)\left(\frac{dy}{dx}\right)^{2}-q(x)y^{2}\right]dx}
{\displaystyle\int_{a}^{b} w(x)y^{2}\,dx}.
\end{equation}

The minimum value of $\mathcal{F}[y]$ corresponds to the lowest eigenvalue of the
system, and the minimizing function is the associated fundamental
eigenfunction. Unlike analytical solutions of Sturm--Liouville problems, where
the overall amplitude of the eigenfunction remains an arbitrary integration
constant that must be fixed by manual normalization, the variational approach
automatically enforces normalization through the denominator of the Rayleigh
quotient during the optimization process.

From a computational perspective, the procedure is straightforward: a trial
function $y_{0}(x)$ satisfying the boundary conditions is initialized, and the
functional $\mathcal{F}[y]$ is minimized using gradient-based optimization to
recover the eigenvalue--eigenfunction pair.

Eigenvalue problems of this form are ubiquitous in quantum mechanics and also arise naturally in classical vibrational systems. To demonstrate the versatility of the proposed variational eigen-solver, we apply it to four representative physical systems: a classical vibrating string, the hydrogen atom, the helium atom, and a proton bound in the $^{28}\mathrm{Si}$ nucleus, modeled using a Woods--Saxon
potential with spin--orbit coupling. 
\subsubsection*{Vibrating String}
We first consider the classical vibrating string, a canonical Sturm–Liouville problem that admits a well-defined variational formulation. The equilibrium mode shapes are obtained by minimizing the elastic energy functional subject to fixed-end boundary conditions. Upon discretizing the functional and applying a gradient-based optimization algorithm, we recover the fundamental normal mode and its associated eigenfrequency (see Table ~\ref{tab:result1} and Figure~\ref{fig:Ritz_fig1}(a)). The numerically computed eigenvalue agrees closely with the analytical result, thereby validating the forward variational implementation.\\ Refer Chapter-3 Section 3.1 of \emph{Supplementary Material} for source codes.
\subsubsection*{Hydrogen Atom}
As a quantum-mechanical example with an exact solution, we apply the framework
to the \emph{hydrogen atom}. The radial Schr\"{o}dinger equation is cast into Sturm--Liouville form with a Coulomb potential and solved by minimizing the Rayleigh quotient. The resulting ground-state energy and wavefunction closely reproduce the known analytical solution, providing a benchmark for the numerical method (see Table ~\ref{tab:result1} and Figure~\ref{fig:Ritz_fig1}(b)). \\ Refer Chapter-3 Section 3.2 of \emph{Supplementary Material} for source codes.

\begin{figure}[!h]
\centering
\includegraphics[width=\linewidth]{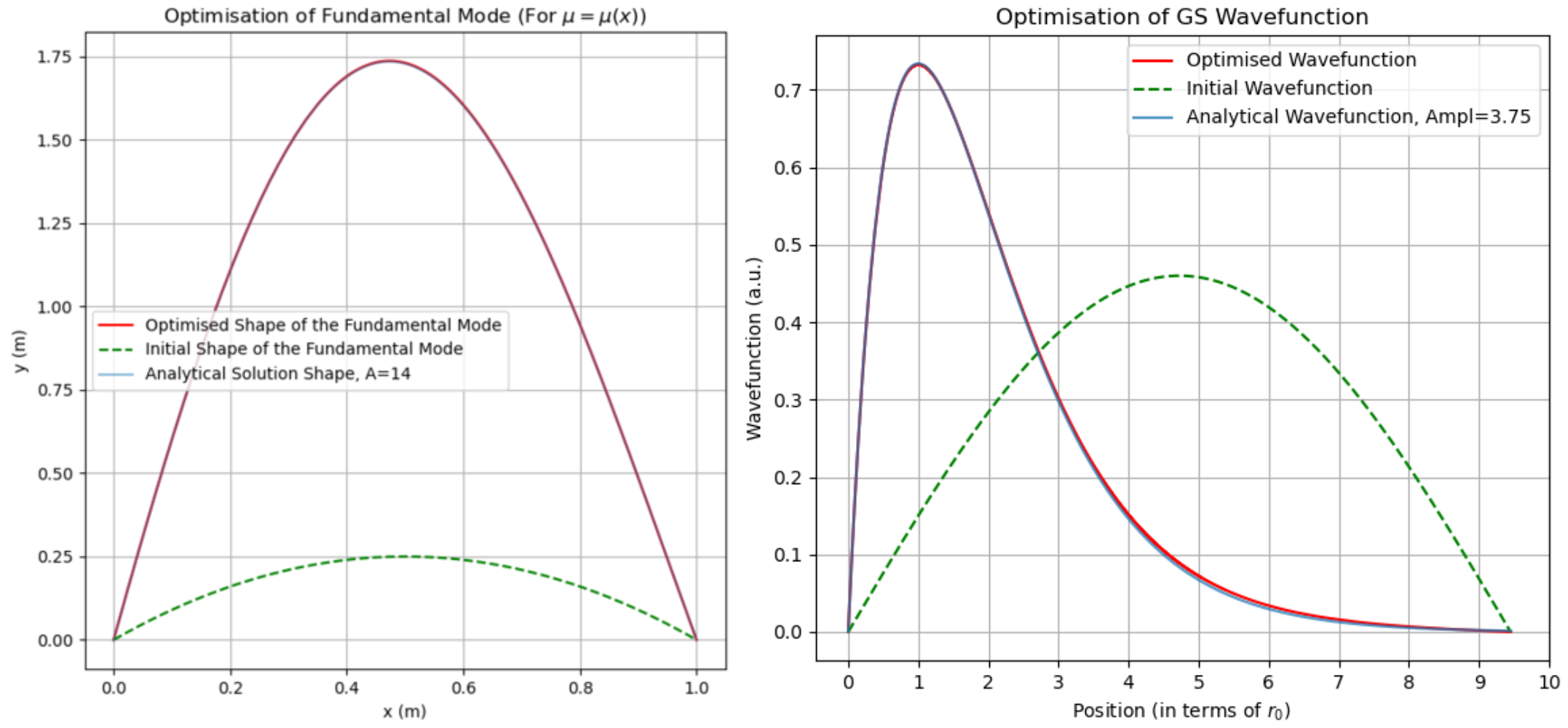} 
\caption{\textbf{Rayleigh-Ritz optimization for (a) a vibrating string and (b) the Hydrogen ground state}. The trial profile (green dashed) relaxes into the final eigen-state (red solid), overlapping perfectly with the analytical benchmark (blue). Spatial coordinates are normalized to $L=1$ and the Bohr radius $a_0$, respectively.}
\label{fig:Ritz_fig1}
\end{figure}

\begin{table*}
    \caption{Summarizes the result of a classical and quantum mechanical system. The computed eigenvalues, which represent the fundamental mode frequency in the vibrational system and the ground state energy in H-atom, match with the theoretical values with a very good accuracy.}
    \renewcommand{\arraystretch}{1.5}
    \begin{ruledtabular}
        \begin{tabular}{ccccc}
           System & Functional Coefficients\footnotemark[1] & Computed Eigenvalue & Theoretical Eigenvalue & Error in Estimation \\ [2pt]  
           \hline
           Classical Vibrating String & \multirow{2}{*}{\shortstack[c]{$p(x)=\mu(x)$\footnotemark[2]; $w(x)=\mu(x)$;\\ $q(x)=0$; $\lambda=\omega_0^2$ (in $rad^2\,s^{-2}$)}} & $\lambda_{comp}\simeq9.7989$ & $\lambda_{theo}\approx9.7987$\footnotemark[3] & $0.0021\%$\\
            & & & &\\
            H-Atom (1-D Radial SWE) & \multirow{2}{*}{\shortstack[c]{$p(r)=-\hbar^2/2m$; $w(r)=-1$;\\ $q(r)=V(r)$\footnotemark[4]; $\lambda=E_{0}$ (in eV)}} & $\lambda_{comp}\simeq-13.609$ & $\lambda_{theo}\approx-13.60$\cite{eisberg1985quantum} & $0.0662\%$\\
            & & & & \\
        \end{tabular}
    \end{ruledtabular}
    \footnotetext[1]{The value of the functional coefficients is obtained by comparing the form of the system's differential equation with Eq.~(\ref{eqn:SLDE})}
    \footnotetext[2]{$\mu(x)=\mu_0+(x-L/2)\Delta$ is the Linear Mass Density of the string of length $L=1$, where $\mu_0=0.954\,gm^{-1}$ and $\Delta=0.5\,gm^{-2}$}
    \footnotetext[3]{See chapter-3 of the supplementary material for the derivation of this value}
    \footnotetext[4]{$V(r)=-\dfrac{e^2}{4\pi\varepsilon_0r}$ is the central coulombic potential}
    \label{tab:result1}
\end{table*}
\subsubsection*{Helium Atom}
The \emph{helium atom} introduces an additional layer of complexity through the
electron--electron repulsion term,
$e^{2}/4\pi\varepsilon_{0} r_{12}$, which precludes an exact analytical solution
of the Schrödinger equation. To make the problem tractable, we adopt a
variational approximation based on two physically motivated assumptions:
\begin{enumerate}
\item \textbf{Screening:} The mutual Coulomb repulsion between the electrons is
modeled as a shielding effect, such that each electron experiences a reduced
effective nuclear charge $Z_{\mathrm{eff}}$ (with $1<Z_{\mathrm{eff}}<2$) instead
of the full nuclear charge.
\item \textbf{Separability:} By replacing the explicit electron--electron
interaction with an effective one-body potential, the electrons are treated as
independent, rendering the Hamiltonian separable.
\end{enumerate}
Within our computational framework, this approximation maps the problem onto a
Sturm--Liouville eigenvalue equation with non-dimensional parameters
$p(r)=-\hbar^{2}/2m$, $w(r)=-1$, and an effective potential
$q(r)=-Z_{\mathrm{eff}}\,e^{2}/4\pi\varepsilon_{0}r$.
\begin{figure}[!h]
    \centering
    \includegraphics[width=\linewidth]{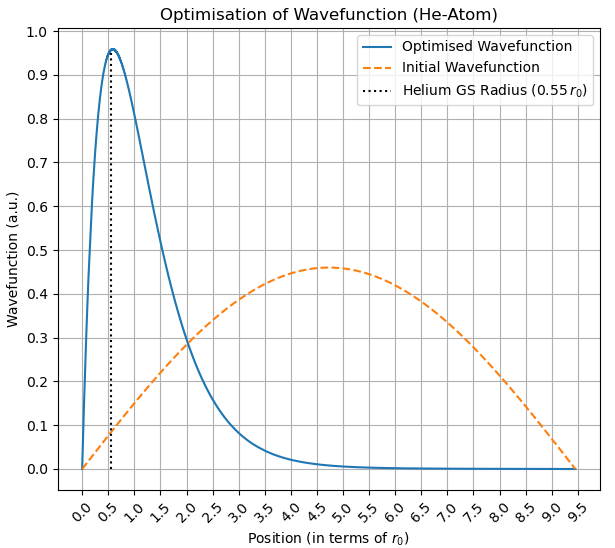}
    \caption{\textbf{Inverse optimization of the Helium radial wavefunction}. The single-electron profile relaxes from the trial state (orange dashed) to the physical ground state (blue solid) as $Z_{\text{eff}}$ is optimized. The radial peak aligns with the theoretical mean radius of $\approx 0.55 a_0$ (vertical dotted line).}
    \label{fig:heatom}
\end{figure}

Unlike standard forward variational calculations, we formulate the helium atom
as an \emph{inverse variational problem}. Using the experimentally measured
ground-state energy\cite{pekeris1958ground} of $-79.0$~eV as the target value, we optimize the parameter
$Z_{\mathrm{eff}}$ so that the computed lowest eigenvalue matches this energy.
The optimization converges to $Z_{\mathrm{eff}}=1.7037$, in excellent agreement
with theoretical benchmarks, deviating by less than $1\%$ from the
Hartree--Fock estimate\cite{griffiths2018introduction} of $Z_{\mathrm{eff}}=1.6875$. The corresponding computed ground-state energy is $-78.99$~eV, and the optimized radial wavefunction is shown in Fig.~\ref{fig:heatom}.\\ Refer Chapter-3 Section 3.3 of \emph{Supplementary Material} for source codes.
\subsubsection*{Silicon Nucleus}
As a final validation of the eigensolver, we compute the proton hole states of the \emph{silicon nucleus}, 
$^{28}\mathrm{Si}$ ($Z=14$), which provides a stringent test of the treatment of the spin--orbit interaction ($V_{ls}$) owing to its closed $1d_{5/2}$ subshell. 

The single-particle motion is described using the combined effective potential
\begin{equation}
V(r)=V_{\text{WS}}(r)+V_{\text{SO}}(r)+V_{\text{CF}}(r)+V_{\text{C}}(r),
\label{Eqn:WS}
\end{equation}
where
\begin{equation}
V_{\text{WS}}(r)=\frac{-V_0}{1+\exp\!\left(\frac{r-R}{a}\right)}
\end{equation}
is the \emph{Woods--Saxon potential}, a schematic mean-field model that captures
nuclear saturation, surface diffuseness, and finite binding through the
phenomenological parameters \(V_0\) (potential well depth), \(a\) (surface
diffuseness), and \(R = R_0 A^{1/3}\) (nuclear radius), where \(A\) is the mass
number and \(R_0 \approx 1.2~\mathrm{fm}\) is the empirical nuclear radius
constant.
\begin{equation}
V_{\text{SO}}(r)=V_{ls}\left(\frac{1}{r}\frac{dV_{\text{WS}}}{dr}\right)
\langle \vec{L}\cdot\vec{S} \rangle
\end{equation}
is the \emph{spin--orbit coupling potential} of strength \(V_{ls}\), and
\begin{equation}
V_{\text{CF}}(r)=\frac{\hbar^2}{2m}\frac{l(l+1)}{r^2}
\end{equation}
is the \emph{centrifugal potential} for the \(l\)th partial wave. The first three
terms in Eq.~(\ref{Eqn:WS}) act on both protons and neutrons. The final term is
the repulsive Coulomb potential experienced only by protons, given by
\begin{equation}
V_{\text{C}}(r)=
\begin{cases}
\dfrac{Ze^2}{2R} \left( 3 - \dfrac{r^2}{R^2} \right), & r \le R, \\[6pt]
\dfrac{Ze^2}{r}, & r > R .
\end{cases}
\end{equation}
\noindent
To obtain the shell energies, we first suppress the spin--orbit term
($V_{ls}=0$) to isolate the central Woods--Saxon potential, thereby obtaining
the geometric centroids of the $1s$, $1p$, and $1d$ shells
(Fig.~\ref{fig:shellsplit}, left). Upon restoring the spin--orbit interaction,
the solver correctly reproduces the characteristic lifting of degeneracy within
each orbital angular momentum manifold (Fig.~\ref{fig:shellsplit}, right).

\begin{figure}
    \centering
    \includegraphics[width=\linewidth]{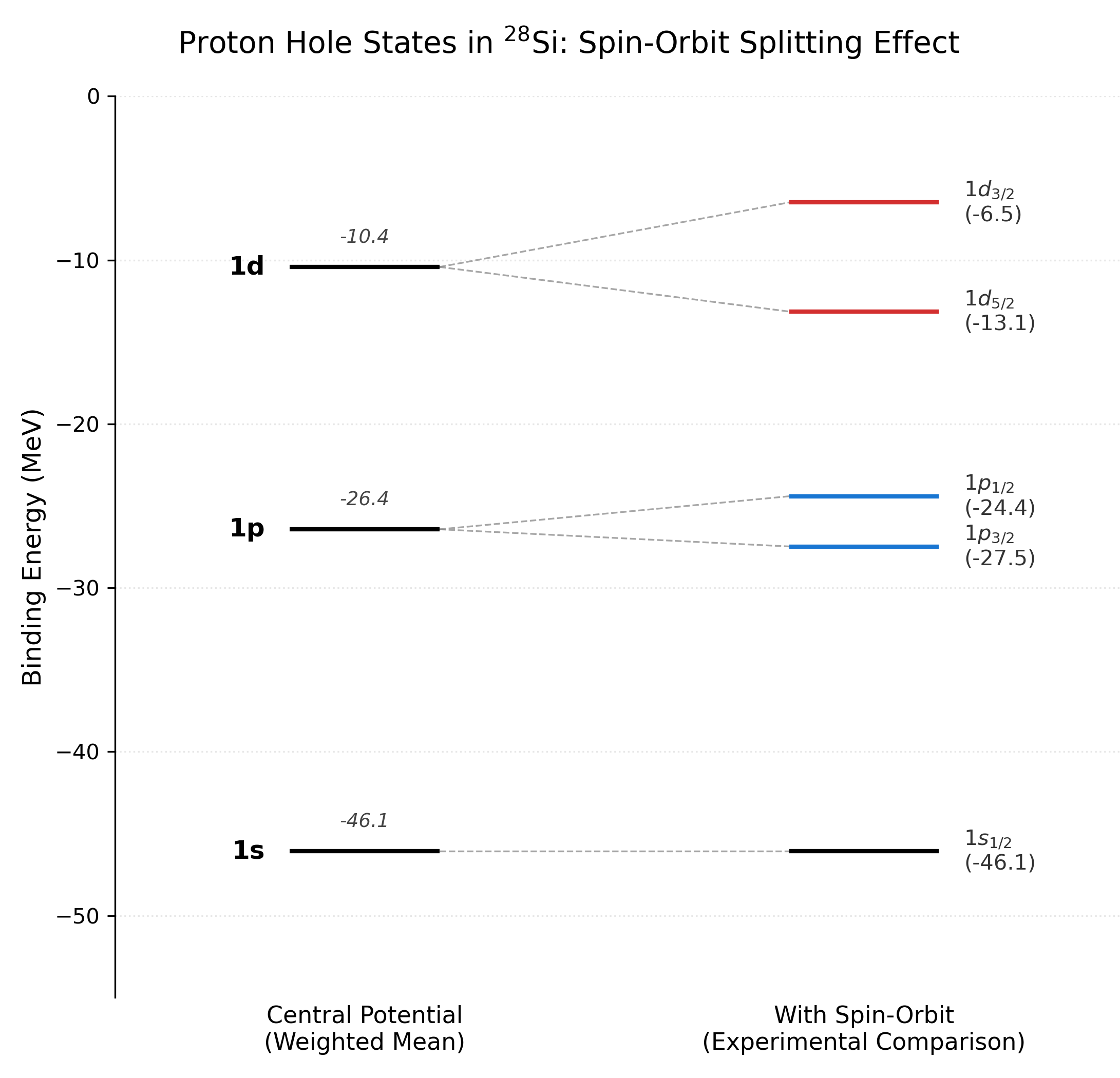}
    \caption{Computed proton energy levels for $^{28}\text{Si}$. (Left) Geometric centroids calculated with a central Woods-Saxon potential ($V_{ls}=0$). (Right) The inclusion of spin-orbit coupling lifts the degeneracy, splitting the $1d$ and $1p$ shells. The visualization highlights the larger splitting for higher orbital angular momentum ($l=2$ vs. $l=1$).}
    \label{fig:shellsplit}
\end{figure}

The computed splittings exhibit the expected angular-momentum dependence: the
$1d$ shell splitting ($\Delta E\!\approx\!6.7$~MeV) exceeds that of the $1p$
shell ($\Delta E\!\approx\!3.1$~MeV), consistent with $(2l+1)$ scaling.
Table~\ref{tab:si28_levels} compares the calculated energies with quasi-free
$(e,e'p)$ data. The model accurately reproduces the deeply bound $1s$ state
near $-46$~MeV and the valence $1d_{5/2}$ state at $-13.14$~MeV; the latter is
slightly overbound relative to experiment ($-11.58$~MeV), a known limitation of
global parameter sets. Overall, the results confirm the solver’s ability to
capture both shell structure and spin--orbit effects in medium-mass nuclei.\\ Refer Chapter-3 Section 3.4 of \emph{Supplementary Material} for source codes.
\begin{table}[h!] 
\centering
\caption{Computed $^{28}\text{Si}$ proton binding energies (MeV). The ``Central'' column reports the geometric centroid ($V_{ls}=0$), while the ``Spin-Orbit'' column includes the $L \cdot S$ coupling. Experimental data are from Ref.~\onlinecite{mougey1976}.}
\label{tab:si28_levels}
\setlength{\tabcolsep}{4pt} 
\renewcommand{\arraystretch}{1.2}
\begin{ruledtabular}
\begin{tabular}{lccc}
\textbf{State} & \textbf{Central} & \textbf{With} & \textbf{Exp.} \\
($n l_j$) & ($V_{ls}=0$) & \textbf{Spin-Orbit} & (e,e'p) \\
\hline
$1s_{1/2}$ & $-46.07$ & $-46.07$ & $\approx -50$ \\

$1p_{3/2}$ & \multirow{2}{*}{$-26.43$} & $-27.48$ & $-35 \pm 4$ \\
$1p_{1/2}$ & & $-24.41$ & $-25 \pm 3$ \\

$1d_{5/2}$ & \multirow{2}{*}{$-10.42$} & $-13.14$ & $-11.58$ \\
$1d_{3/2}$ & & $-6.48$ & -- \\
\end{tabular}
\end{ruledtabular}
\end{table}
\subsubsection{Boundary Value Problem: The Dirichlet's Principle}
A physically transparent illustration of \emph{Dirichlet’s principle} arises in
steady--state heat conduction. When the temperature field $T(\mathbf{r})$ has
reached equilibrium, it becomes time--independent and is governed by a balance
between thermal diffusion and internal heat sources. Starting from Fourier’s
law and energy conservation, the governing equation can be written as
\begin{equation}
\nabla \cdot \left( k \nabla T \right) + q(T) = 0,
\label{Eqn:heatbal}
\end{equation}
where $k$ is the thermal conductivity and $q(T)$ denotes the volumetric rate of
internal heat generation.

According to Dirichlet’s principle, the solution of this boundary value problem
is the temperature field that minimizes the energy functional
\begin{equation}
\mathcal{E}[T]
=
\int_{\Omega}
\left[
\frac{1}{2}\,k\,|\nabla T|^{2}
-
Q(T)
\right] dV,
\end{equation}
where the potential $Q(T)$ satisfies $dQ/dT = q(T)$. Requiring the first
variation of $\mathcal{E}[T]$ to vanish, $\delta \mathcal{E}[T]=0$, subject to
the prescribed boundary conditions, yields the Euler--Lagrange equation, which
recovers the steady--state heat conduction equation above.

This variational formulation allows steady--state heat conduction problems to
be treated as optimization problems: the physically realized temperature
distribution corresponds to the configuration of minimum energy consistent
with the imposed boundary conditions. We solve the following representative boundary value problems to illustrate the given idea.

\subsubsection*{The Homogeneous Square Plate: A Benchmark of Accuracy}

For a homogeneous square plate with uniform thermal conductivity $\kappa$ and
no internal heat generation, the steady--state temperature field satisfies
\emph{Laplace’s equation}. In two dimensions, the governing equation takes the form
\begin{equation}
\nabla \cdot \left( \kappa \nabla T \right) = 0,
\end{equation}
which reduces to
\begin{equation}
\nabla^{2} T = 0
\end{equation}
for constant $\kappa$. This equation describes pure diffusive heat transport
driven solely by the imposed boundary conditions. This problem provides a
standard benchmark for validating numerical methods for solving elliptic
partial differential equations. We apply our variational optimization
framework to this problem to assess both its accuracy and robustness.

Starting from a randomly initialized interior temperature field consistent
with the prescribed boundary conditions, the optimizer iteratively minimizes
the discretized Dirichlet energy functional and converges to its ground state.
The resulting temperature distribution is validated by comparison with a
solution obtained using a traditional \emph{Jacobi iterative solver}\cite{chapra2011numerical} as shown in Fig.~\ref{fig:2D_V}. Figure~\ref{fig:2D_E} depicts that difference between the two solutions is minimal, with the largest deviations occurring near the center of the plate where temperature gradients are weakest. This confirms that the variational and iterative approaches yield nearly identical steady--state solutions. \\ Refer Chapter-4 Section 4.1 of \emph{Supplementary Material} for source codes.
\begin{figure}[!h]
\centering
    \begin{subfigure}[b]{0.50\linewidth}
    \centering
        \includegraphics[width=\linewidth]{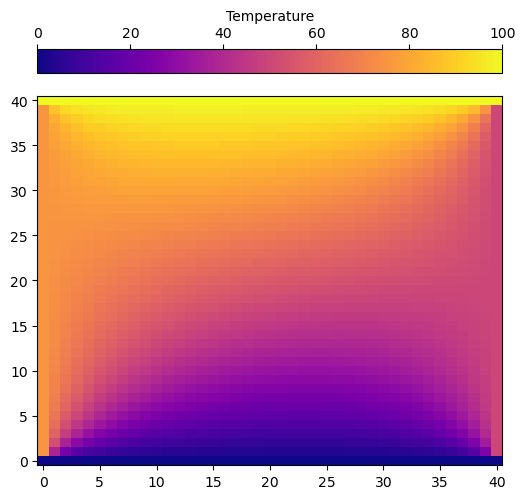}
        \caption{}
        \label{fig:2D_V}
    \end{subfigure}
    \begin{subfigure}[b]{0.49\linewidth}
    \centering
        \includegraphics[width=\linewidth]{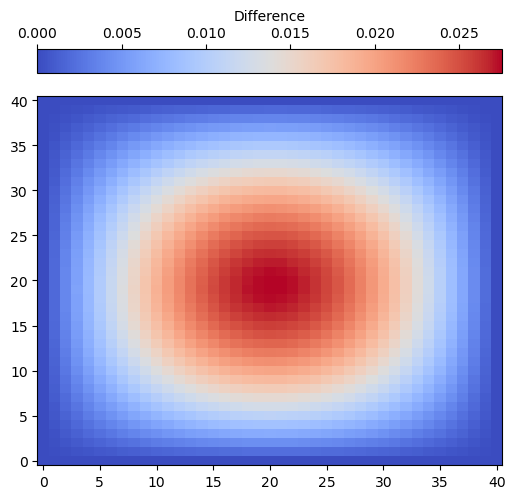} 
        \caption{}
        \label{fig:2D_E}
    \end{subfigure}
    \caption{\textbf{Simulation of the temperature distribution} on a 41 x 41 grid square plate with boundary conditions $u(x,0)=0$, $u(x,40)=100$, $u(0,y)=75$, and $u(40,y)=50$. \textbf{(a)} The temperature distribution obtained using a variational solver. \textbf{(b)} The absolute difference between the grid values obtained from the variational solver and an iterative solver.}
    \label{fig:squareplate}
\end{figure}
\subsubsection*{Inhomogeneous Rod: Extension to Variable Coefficients and Sources}
To demonstrate the extensibility of the variational approach, we next consider
a one--dimensional rod with spatially varying thermal conductivity and internal
heat generation. When the conductivity $\kappa(x)$ varies along the rod and a
source term $q(x)$ is present, Eq.~\ref{Eqn:heatbal} takes the inhomogeneous
\emph{Poisson} form
\begin{equation}
\frac{d}{dx}
\left(
\kappa(x)\frac{dT}{dx}
\right)
+
q(x)
=
0.
\end{equation}
Direct finite--difference solutions become cumbersome in this case because the
differential operator acts on the product of $\kappa(x)$ and the temperature
gradient. In contrast, the variational formulation naturally incorporates this
complexity by minimizing the functional
\begin{equation}
E[T]
=
\int_{0}^{L}
\left[
\frac{1}{2}\,\kappa(x)
\left(\frac{dT}{dx}\right)^{2}
-
q(x)\,T
\right] dx.
\end{equation}
Upon discretization, the functional becomes
\begin{equation}
E_h[T]
=
\sum_i
\left[
\frac{1}{2}\,\kappa(x_i)
\left(
\frac{T_{i+1}-T_i}{\Delta x}
\right)^2
-
q(x_i)\,T_i
\right]\Delta x.
\end{equation}
Minimization of $E_h[T]$ using gradient--based optimization yields the
steady--state temperature distribution along the rod, consistent with the
imposed boundary conditions as shown in Fig.~\ref{fig:inhomo_rod}. \\ Refer Chapter-4 Section 4.2 of \emph{Supplementary Material} for source codes.
\begin{figure}[!h]
\centering
    \begin{subfigure}[b]{0.51\linewidth}
        \includegraphics[width=\linewidth]{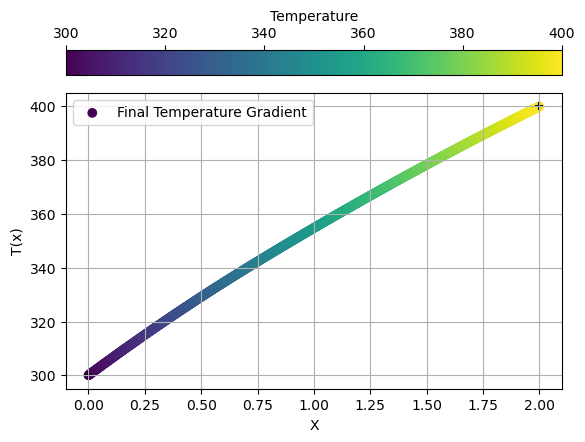}
        \caption{Temperature distribution along the rod obtained using the variational solver.}
        \label{fig:R_V}
    \end{subfigure}
    \hfill
    \begin{subfigure}[b]{0.48\linewidth}
        \includegraphics[width=\linewidth]{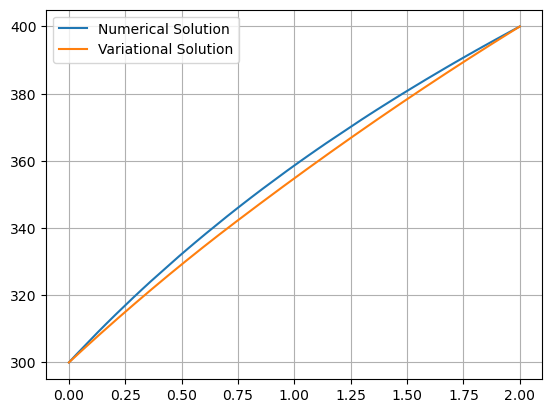}
        \caption{Comparison between the variational solution and a standard numerical solution.}
        \label{fig:R_C}
    \end{subfigure}
    \caption{Heat conduction in a rod of length $L=2$ with fixed boundary
    temperatures $T(0)=300$ and $T(L)=400$. The thermal conductivity varies
    linearly as $\kappa(x)=\kappa_0(1+\beta x/L)$, and the internal heat source
    is $q(x)=q_0\sin(\pi x/L)$. The close agreement between the variational and
    numerical solutions demonstrates the robustness of the optimization-based
    approach.}
    \label{fig:inhomo_rod}
\end{figure}

\subsubsection*{Nonlinear Differential Equation}

Thus far, we have restricted attention to linear differential equations.
However, the variational framework developed here is not limited to linear
systems and can be applied equally naturally to nonlinear boundary value
problems. As a representative example, we consider the nonlinear equation
\begin{equation}
\frac{d^{2}y}{dx^{2}} = y^{3} + y,
\end{equation}
subject to appropriate boundary conditions.

Rather than solving this equation directly, we interpret it as the
Euler--Lagrange equation associated with the minimization of the functional
\begin{equation}
E[y]
=
\int_{a}^{b}
\left[
\frac{1}{2}\left(\frac{dy}{dx}\right)^{2}
-
\left(
\frac{1}{4}y^{4}
+
\frac{1}{2}y^{2}
\right)
\right] dx.
\end{equation}
The steady--state solution corresponds to the function $y(x)$ that minimizes
$E[y]$ while satisfying the prescribed boundary conditions.

Upon discretization on a uniform grid with spacing $\Delta x$, the functional
reduces to
\begin{equation}
E_h[y]
=
\sum_i
\left[
\frac{1}{2}
\left(
\frac{y_{i+1}-y_i}{\Delta x}
\right)^{2}
-
\left(
\frac{1}{4}y_i^{4}
+
\frac{1}{2}y_i^{2}
\right)
\right]\Delta x,
\end{equation}
thereby transforming the nonlinear boundary value problem into a
finite--dimensional optimization problem. Figure~\ref{fig:Nonlinear} compares the solution obtained using the optimization approach with the analytical solution.
\begin{figure}
\centering
\includegraphics[width=\linewidth]{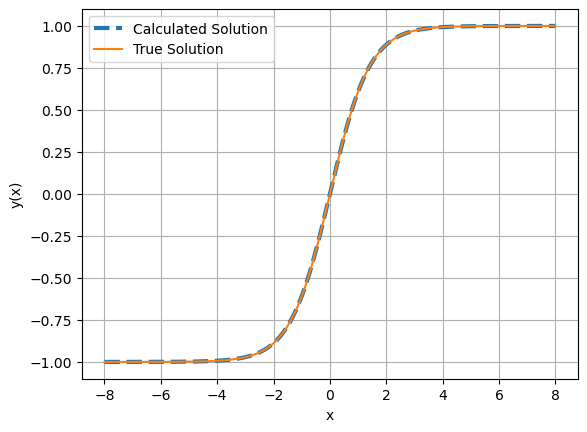}
\caption{For solving this problem we assumed the boundary conditions to be \texttt{y(-L)=-1 and y(L)=1} and we can see how our solution matches perfectly with the analytical solution , $y(x)=tanh(\frac{x}{\sqrt{2}})$}
\label{fig:Nonlinear}
\end{figure}

From a physical perspective, this model may be interpreted as steady--state
heat conduction with nonlinear internal heat generation, where the equilibrium
temperature profile emerges as the configuration that optimally balances
diffusion against nonlinear source effects by minimizing an underlying energy
functional.  \\ Refer Chapter-4 Section 4.3 of \emph{Supplementary Material} for source codes.

\subsubsection{Non-Conservative Systems: The Residual Loss Functional}
\begin{figure*}
\centering
    \begin{subfigure}[b]{0.45\linewidth}
        \includegraphics[width=\linewidth]{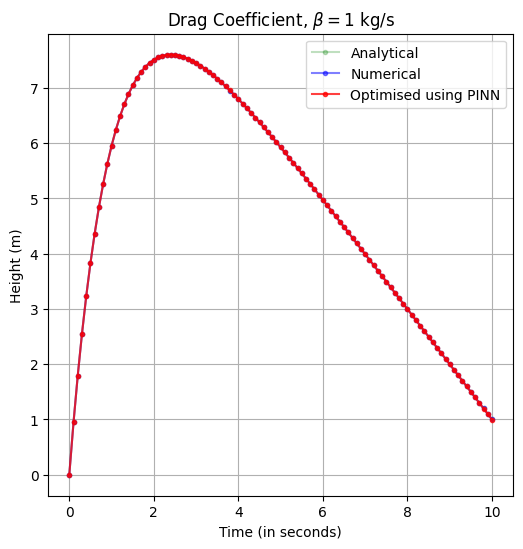}
        \caption{}
        \label{fig:proj_drag}
    \end{subfigure}
    \hfill
    \begin{subfigure}[b]{0.49\linewidth}
        \includegraphics[width=\linewidth]{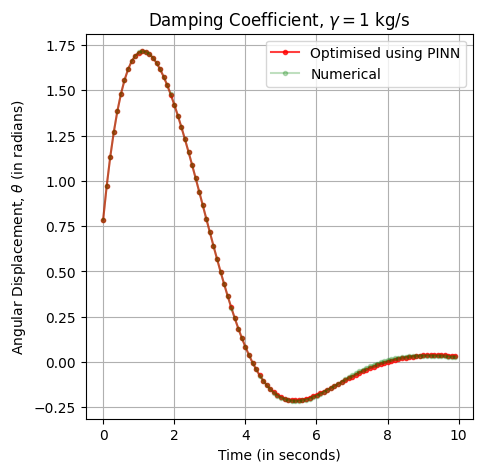} 
        \caption{}
        \label{fig:pend_drag}
    \end{subfigure}
    \caption{\textbf{Forward modeling of non-conservative systems using Physics-Informed Neural Networks (PINNs)}. The network reconstructs the time evolution for \textbf{(a)} a projectile subject to linear air drag and \textbf{(b)} a damped simple pendulum. In both cases, the optimized PINN predictions (red/orange) show excellent agreement with the numerical benchmarks (green/dashed), demonstrating the solver's ability to capture dissipative dynamics without direct supervision.}
    \label{fig:pinn_res}
\end{figure*}
Since dissipative systems do not possess a conserved quantity such as energy or action that can be extremized, variational formulations based on minimization principles are generally unavailable. To address this limitation, we employ \emph{Physics-Informed Neural Networks (PINNs)}, in which the solution is obtained by minimizing a functional constructed from the squared residual of the governing differential equation. In this framework, the differential equation itself acts as a soft constraint that guides the learning process.

We demonstrate the versatility of this approach by applying the PINN solver to two representative \textbf{non-conservative dynamical systems} in which mechanical energy is continuously dissipated due to friction. In both cases, the framework is used not only for \emph{forward trajectory prediction} but also for \emph{inverse parameter estimation}.

\subsubsection*{Projectile Motion with Linear Drag} 
    The vertical motion of a particle subject to gravity and linear air resistance is governed by\cite{taylor2005classical}
    \begin{equation}
        m \ddot{y} + \beta \dot{y} + mg = 0 .
    \end{equation}
    Here, the PINN is trained using sparse trajectory data, while treating the drag coefficient $\beta$ and the initial conditions as trainable parameters. As shown in Fig.~\ref{fig:proj_drag}, the network successfully reconstructs the asymmetric trajectory characteristic of dissipative motion. In the corresponding inverse problem (Table~\ref{tab:inv_model}), the learned drag coefficient closely matches the true value, recovering $\beta_{\text{actual}} = 1.0~\text{kg/s}$ with high precision $\beta_{\text{opt}} \approx 1.0015~\text{kg/s}$. \\ Refer Chapter-5 Section 5.1 of \emph{Supplementary Material} for source codes.

\subsubsection*{Damped Pendulum}  
    We next consider a nonlinear dissipative system—a pendulum of length $l$ subject to viscous damping—described by\cite{taylor2005classical}
    \begin{equation}
        ml \ddot{\theta} + \gamma l \dot{\theta} + mg \sin \theta = 0 .
    \end{equation}
    This example tests the solver’s robustness in handling nonlinear restoring forces combined with velocity-dependent dissipation. The PINN accurately captures the progressive decay of the oscillation amplitude, as illustrated in Fig.~\ref{fig:pend_drag}. In the inverse modeling task (Table~\ref{tab:inv_model}), the network converges to the correct damping coefficient ($\gamma \approx 1.007~\text{kg/s}$) and initial angular displacement ($\theta_0 \approx 0.785~\text{rad}$), starting from randomized initial guesses. \\ Refer Chapter-5 Section 5.2 of \emph{Supplementary Material} for source codes

\begin{table}[h]
\centering
\caption{Inverse modeling of non-conservative systems. The PINN accurately recovered hidden physical parameters (friction coefficients and initial conditions) from sparse trajectory data.}
\label{tab:inv_model}
\setlength{\tabcolsep}{5pt}
\renewcommand{\arraystretch}{1.3}
\begin{ruledtabular}
\begin{tabular}{l l c c c}
\textbf{System} & \textbf{Parameter} & \textbf{Actual} & \textbf{Optimized} & \textbf{Abs.} \\
 & & & & \textbf{Error}\\
\hline
\multirow{3}{*}{\shortstack[l]{Projectile\\with Drag}} 
 & Drag $\beta$ (kg/s) & $1.0000$ & $1.0015$ & $0.0015$ \\
 & $y(0)$ (m) & $0.0000$ & $0.0001$ & $0.0001$ \\
 & $\dot{y}(0)$ (m/s) & $10.0000$ & $10.0062$ & $0.0062$ \\
\hline
\multirow{3}{*}{\shortstack[l]{Damped\\Pendulum}} 
 & Damping $\gamma$ (kg/s) & $1.0000$ & $1.0071$ & $0.0071$ \\
 & $\theta(0)$ (rad) & $0.7854$ & $0.7857$ & $0.0002$ \\
 & $\omega(0)$ (rad/s) & $2.0000$ & $2.0049$ & $0.0049$ \\
\end{tabular}
\end{ruledtabular}
\end{table}

\section{Discussion and Scope}

The primary strength of the framework presented here lies in its conceptual unification. We have shown that a single algorithmic strategy—the minimization of a scalar functional—can be applied consistently across a wide range of physical systems. Whether deriving Snell’s law from Fermat’s principle, computing the chaotic trajectory of a double pendulum via the principle of least action, or determining the eigen-spectrum of the \(^{28}\mathrm{Si}\) nucleus using the Ritz--Rayleigh method, the underlying computational engine remains essentially unchanged. This versatility across both linear and nonlinear systems enables students to view these problems not as isolated mathematical exercises, but as diverse manifestations of a common optimization principle that permeates physics.

At the same time, it is essential to delineate the mathematical boundaries of the traditional variational approach. For a system of differential equations to be derivable from a stationary principle—namely, for a Lagrangian functional to exist—the system must satisfy the Helmholtz conditions, which define the inverse problem of the calculus of variations \cite{nigam2016briefreviewhelmholtzconditions}. Physically, these conditions restrict standard Lagrangian mechanics to conservative systems governed by self-adjoint operators. Dissipative phenomena, such as air drag in projectile motion or friction in a damped pendulum, introduce odd-order time derivatives (e.g., \(\gamma \dot{x}\)) that violate these criteria. As a result, such systems cannot be treated by minimizing a physical action, since no corresponding scalar potential exists within the conventional formalism.

This limitation motivates the introduction of Physics-Informed Neural Networks (PINNs) as a natural generalization of the variational paradigm. By minimizing the squared residual of the governing differential equation, PINNs replace the action functional with a well-defined loss function that does not require symplectic structure or self-adjointness. The only essential requirement is the continuity of the differential equation itself. This allows the optimization-based workflow to be extended seamlessly to non-conservative and dissipative systems. In this sense, while traditional variational principles underpin the fundamental interactions of nature—classical mechanics, electromagnetism, and quantum theory—PINNs complete the pedagogical toolkit by enabling variational thinking to be applied to phenomenological systems characterized by friction, damping, and decay.

To facilitate adoption in instructional settings, all numerical examples presented in this work are implemented using open-source Python tools, and the fully documented Python codes are provided in the supplementary material. These resources are expected to enable instructors to reproduce the results, adapt the workflows to their own courses, and implement the proposed pedagogical approach with minimal overhead, thereby supporting reproducible and computation-driven instruction in undergraduate physics.

\section{Pedagogical Remarks}

The numerical solvers developed in this work are intended to bridge the gap between analytical theory and computational practice. Rather than treating computation as a post-processing tool, the proposed framework integrates numerical optimization directly into the formulation of physical problems, allowing students to engage with variational principles in an explicit and constructive manner. We envision three primary modes through which this material can be incorporated into the undergraduate physics curriculum.

\begin{enumerate}
    \item \textbf{A Unified Computational Framework:}  
    Undergraduate physics courses often compartmentalize topics such as classical mechanics, optics, and quantum theory into distinct subject areas. By employing a common solver framework to compute the eigenfrequency of a vibrating string, determine least-action trajectories in mechanics, and evaluate nuclear shell structure in \(^{28}\mathrm{Si}\), students are exposed to the underlying mathematical universality of physics. This approach encourages students to recognize variational principles as unifying algorithmic ideas rather than topic-specific derivations.

    \item \textbf{Introduction to Scientific Machine Learning:}  
    The Physics-Informed Neural Network (PINN) examples presented here provide an accessible and physics-driven entry point to machine learning. Unlike conventional data-fitting exercises, inverse modeling of systems such as the damped pendulum demonstrates how physical laws—encoded as differential equation residuals within the loss function—act as regularization constraints. This enables students to use machine learning not only for prediction, but also for parameter discovery in non-conservative systems where traditional variational methods fail.

    \item \textbf{Suggested Student Projects:}
\begin{enumerate}
    \item \emph{Visualizing the Calculus of Variations:}  
    Students can animate the iterative relaxation of an initial randomized trial path into a physical trajectory by minimizing the classical action. This process develops geometric intuition for the Euler--Lagrange equations that static analytical derivations often obscure. As a canonical benchmark, the same framework can be used to rediscover the brachistochrone (cycloid) curve by minimizing the time of descent.

    \item \emph{A Universal Eigenvalue Solver:}  
    Using the Ritz--Rayleigh variational framework, students can solve eigenvalue problems by minimizing appropriate energy-based quantities. In classical systems, this includes minimizing elastic or strain energy to determine natural frequencies of vibrating membranes and beams, as well as minimizing total potential energy to identify critical buckling loads in column like structures. In quantum mechanics, the approach minimizes the expectation value of the Hamiltonian to obtain bound-state energies\cite{schroeder2017variational}. Extension to nuclear structure is achieved by minimizing the Woods--Saxon energy to explore shell evolution and the emergence of magic numbers in \(^{28}\mathrm{Si}\).

    \item \emph{Physics-Informed Discovery (PINNs):}  
    For systems that do not admit a physical action principle, students may instead minimize a physics-informed loss function defined by the squared residual of the governing differential equation. Using noisy synthetic data from a damped oscillator, or experimental measurements more generally, inverse PINN solvers can be employed to infer unknown parameters such as friction coefficients or other relevant system parameters. Advanced extensions include coupled systems, such as the Lotka--Volterra predator--prey model \cite{lotkavolt}, demonstrating how physical constraints guide learning in data-limited regimes.
\end{enumerate}

\end{enumerate}

\section{Conclusion}

This work presents a comprehensive variational toolkit for undergraduate physics, demonstrating that a single minimization-based strategy can be adapted to describe a broad spectrum of physical phenomena. The flexibility of the framework is reflected in the choice of functional appropriate to each domain, ranging from optical path length in refraction and action minimization in dynamical systems to energy functionals in quantum mechanics and steady-state heat conduction. The inclusion of Physics-Informed Neural Networks (PINNs) further extends this toolkit by overcoming the intrinsic limitations of traditional variational mechanics. Whereas classical functionals are constrained by the Helmholtz criteria and fail in the presence of dissipation, PINNs treat the differential equation residual itself as a pseudo-functional, thereby enabling variational reasoning for non-conservative systems and inverse problems. Viewed in this light, modern gradient-based optimization and machine learning are not departures from classical physics, but natural computational descendants of the variational ideas introduced by Fermat, Euler, and Lagrange.

\begin{acknowledgments}
The authors gratefully acknowledge Prof. Kedar Khare, Professor, Optics and Photonics Centre, Indian Institute of Technology Delhi, New Delhi-110016 for conceiving the original idea that motivated this work and for helpful discussions during its early development. Financial support from the Department of Biotechnology (DBT), Government of India, under the DBT Star College Scheme (Grant No. HRD-11011/20/2022-HRD-DBT) is also sincerely acknowledged.
\end{acknowledgments}

\section*{Authors Declaration}

The authors declare that they have no conflicts of interest.

\bibliography{aipsamp}

\end{document}